\documentclass[pra,twocolumn,superscriptaddress,showpacs,footinbib]{revtex4}

\usepackage{amsfonts}
\usepackage{amssymb}
\usepackage{amsmath}
\usepackage{hyperref}
\usepackage{graphicx}
\usepackage{longtable}
\usepackage{bm}

\newcommand{\bff}{\mathbf{F}}
\newcommand{\bfv}{\mathbf{v}}
\newcommand{\bfG}{\mathbf{G}}

\newcommand{\barG}{\overline{G}}
\newcommand{\bardelt}{\overline{\delta t}}
\newcommand{\bardelttwo}{\overline{\delta t^{2}}}
\newcommand{\bardelphi}{\overline{\delta\phi}}
\newcommand{\bardelphitwo}{\overline{\delta\phi^{2}}}
\newcommand{\barom}{\overline{\omega}}

\newcommand{\calG}{\mathcal{G}}
\newcommand{\calO}{\mathcal{O}}
\newcommand{\calA}{\mathcal{A}}
\newcommand{\calI}{\mathcal{I}}
\newcommand{\calGU}{\mathcal{G}_{U}}

\newcommand{\calF}{\mathcal{F}}

\newcommand{\bldc}{{\textbf c}}
\newcommand{\blde}{{\textbf e}}
\newcommand{\bldf}{{\textbf f}}
\newcommand{\bldg}{{\textbf g}}

\newcommand{\bldw}{{\textbf w}}
\newcommand{\bldx}{{\textbf x}}
\newcommand{\bldy}{{\textbf y}}

\newcommand{\bldgam}{\mbox{\boldmath $\gamma$}}
\newcommand{\bldlam}{\mbox{\boldmath $\lambda$}}
\newcommand{\bldb}{\mbox{\boldmath $\beta$}}
\newcommand{\bldsig}{\mbox{\boldmath $\sigma$}}
\newcommand{\bldom}{\mbox{\boldmath $\omega$}}
\newcommand{\bldcdot}{\mbox{\boldmath $\cdot$}}
\newcommand{\bfhatz}{\hat{\mathbf{z}}} 
\newcommand{\bfhatx}{\hat{\mathbf{x}}}
\newcommand{\bfhaty}{\hat{\mathbf{y}}}

\newcommand{\calN}{\mathcal{N}}
\newcommand{\barNf}{\overline{\mathcal{N}_{f}}}

\newcommand{\meanP}{\overline{P}}
\newcommand{\specdenphi}{S_{\phi}(f)}
\newcommand{\specdenN}{S_{N}(f)}
\newcommand{\meann}{\overline{n}}

\newcommand{\sgn}{\mathrm{sgn}}

\begin{document}
\title{Improving Quantum Gate Performance through Neighboring Optimal Control}

\date{\today}

\author{Yuchen Peng}
\affiliation{Department of Physics, University of Maryland,
College Park, MD 20742}
\author{Frank Gaitan}
\affiliation{Laboratory for Physical Sciences, 8050 Greenmead Dr,
College Park, MD 20740}

\begin{abstract}
Successful implementation of a fault-tolerant quantum computation on a system 
of qubits places severe demands on the hardware used to control the many-qubit 
state. It is known that an accuracy threshold $P_{a}$ exists for any quantum 
gate that is to be used in such a computation. Specifically, the error 
probability $P_{e}$ for such a gate must fall below the accuracy threshold: 
$P_{e} < P_{a}$. Estimates of $P_{a}$ vary widely, though $P_{a}\sim 10^{-4}$ 
has emerged as a challenging target for hardware designers. In this paper we 
present a theoretical framework based on neighboring optimal control that takes 
as input a good quantum gate and returns a new gate with better performance. 
We illustrate this approach by applying it to all gates in a universal set of 
quantum gates produced using non-adiabatic rapid passage that has appeared in 
the literature. Performance improvements are substantial, both for ideal and 
non-ideal controls. Under suitable conditions detailed below, all gate error 
probabilities fall well below the target threshold of $10^{-4}$.
\end{abstract}


\maketitle

\section{Introduction}
\label{sec1}

It is now well-established that reliable quantum computing is possible, even
in the presence of decoherence and imperfect control \cite{ft1,ft2,ft3,ft4,ft5,
ft6,ft7,ft8}. In spite of this important result, it is also well-appreciated 
that significant technical obstacles currently stand in the way of building a 
scalable quantum computer. One major challenge is finding a way to implement 
a high-fidelity universal set of quantum gates from which an arbitrary quantum 
computation can be constructed. The accuracy threshold $P_{a}$ provides a 
quantitative measure of the accuracy demanded of a quantum gate. Specifically, 
if a quantum gate is to be used in a reliable quantum computation, the 
probability $P_{e}$ that it produces an error must be less than the accuracy 
threshold: $P_{e} < P_{a}$. The accuracy threshold is a function of the 
quantum error correcting code used to protect the computational data, and the 
fault-tolerant procedures used to control the spread of errors during the 
computation. Estimates of $P_{a}$ vary widely, from as small as $10^{-6}$, to 
as large as a few times $10^{-3}$. Over the years, the value $P_{a}\sim 
10^{-4}$ has emerged as a challenging target for quantum hardware designers.
One of the central problems in quantum control is finding a way to implement 
a universal set of quantum gates whose gate error probabilities are all less 
than $10^{-4}$.

To apply a quantum gate, a control field $\bff (t)$ is applied to a quantum 
system over a time $T$, causing a time-varying unitary transformation 
$U(t)$ to act on the quantum state. When designing a quantum gate, the task
is to find the control field $\bff (t)$ that applies a target gate $U_{tgt}$ to
the quantum state (viz.~$U(t=T) = U_{tgt}$). In optimal control theory, the 
task is to find a control field profile $\bff_{\ast}(t)$ that produces a 
high-fidelity approximation $U(t)$ to the target gate $U_{tgt}$, while 
simultaneously minimizing a cost function that depends on the state $U(t)$ 
and control field $\bff (t)$. The control profile $\bff_{\ast} (t)$ is called 
the optimal control, and the corresponding unitary $U_{\ast}(t)$ is called 
the optimal (state) trajectory. Note that a perturbation of the dynamics can 
cause an optimal trajectory and control to become non-optimal. However, if the 
perturbation is small, the optimal control problem can be linearized about the 
original optimal solution, and a family of perturbed optimal trajectories  
determined from a single feedback control law. In the classical literature
this perturbed control problem is referred to as neighboring optimal 
control \cite{stengel}.

In this paper we consider the problem of making a good quantum gate better. 
It is assumed that we know the control field profile $\bff_{0}(t)$ that produces
a good approximation $U_{0}(t=T)$ to a target gate $U_{tgt}$. We extend the 
strategy of neighboring optimal control to the dynamics of a quantum system and 
use it to determine the control modification $\Delta\bff (t)$ that produces an 
improved approximation $U(t=T)$ to the target $U_{tgt}$. To illustrate the 
general theory, we use it to improve the performance of all gates in a universal
set of quantum gates produced using non-adiabatic rapid passage that has been 
studied in the literature \cite{trp1,trp2,trp3,trp4,trp5,trp6,trp7,trp8}. 
We examine both ideal and non-ideal controls, and show that under 
suitable conditions, all gate error probabilities fall well below the target 
threshold of $10^{-4}$. Although we focus on a target threshold 
$P_{a}= 10^{-4}$ throughout this paper, it is important to note that for 
surface and color quantum error correcting codes, the accuracy threshold 
satisfies $P_{a}\sim 10^{-3}$ 
\cite{homcodes1,homcodes2,homcodes3,homcodes4,homcodes5}. For these codes, 
the neighboring optimal control improved non-adiabatic rapid passage gates 
all operate at least two orders of magnitude below threshold, even for 
non-ideal control.

The structure of this paper is as follows. In Section~\ref{sec2} we lay out the
general theoretical framework for applying neighboring optimal control to the
problem of improving the performance of a good quantum gate. We use the
Schrodinger equation to determine the equation of motion for the gate 
modification $\delta U(t) = U^{-1}_{0}(t)U(t)$ in Section~\ref{sec2a}; 
formulate the cost function for the optimization in Section~\ref{sec2b}; 
derive the system of equations that determine the optimal solution in 
Section~\ref{sec2c}, and present two strategies for obtaining that solution in 
Section~\ref{sec2d}. We illustrate the general method in Section~\ref{sec3} 
by using it to improve the performance  of a universal set of quantum gates. 
In the interests of clarity, Section~\ref{sec3} examines the case of the 
Hadamard gate in detail, with results for the remaining quantum gates 
presented in Appendix~\ref{appendixRemainingResults}. Finally, 
Section~\ref{sec4} summarizes our results; Appendix~\ref{appendixTRP} 
briefly reviews the form of non-adiabatic rapid passage used to produce 
the initial universal set of quantum gates examined in Section~\ref{sec3};
Appendix~\ref{appendixB} derives a formula needed in Section~\ref{sec2d};
and Appendix~\ref{appendixNoiseModel} describes the noise model and simulation
protocol used to examine phase jitter effects in Section~\ref{sec3c2}.

\section{General Theory}
\label{sec2}

In this Section we introduce a general theoretical framework that takes a good 
quantum gate $U_{0}(t)$ as input, and returns a better one $U(t)$. 
Section~\ref{sec2a} determines the equation of motion for the gate 
modification $\delta U(t) = U^{\dagger}_{0}(t)U(t)$; Section~\ref{sec2b} 
constructs the cost function whose minimum determines the optimal gate 
modification; Section~\ref{sec2c} varies the cost function to determine the 
equations that govern the optimization; and Section~\ref{sec2d} presents two 
strategies for obtaining their solution. In Section~\ref{sec3} we illustrate 
the general method by using it to improve the performance of all gates in a 
universal set of quantum gates.

In this paper we follow the standard physics convention of denoting a column 
vector by a boldface symbol $\bfv$; a row vector by the Hermitian adjoint of a 
boldface symbol $\bfv^{\dagger}$; and a matrix by a non-boldface symbol $M$.
Thus $M\bfv$ represents the product of a matrix $M$ with a column vector 
$\bfv$, and $\bldy^{\dagger}\bldx$ is the product of a row vector 
$\bldy^{\dagger}$ with a column vector $\bldx$.

\subsection{Gate modification dynamics}
\label{sec2a}

Consider a Hamiltonian $H(t) = H[\bff (t)]$ that is a functional of a 
control field $\bff (t) = \bff_{0}(t) + \Delta\bff (t)$ that contains a small 
variation $\Delta\bff (t)$ about a nominal control field $\bff_{0} (t)$. 
Expanding the Hamiltonian $H(t)$ about $\bff_{0}(t)$ gives
\begin{eqnarray}
H(t) & = & H[\bff_{0}(t)] + \sum_{j=1}^{3}\left. \frac{\delta H}{\delta F_{j}}
                   \right|_{\bff_{0}}\Delta F_{j} + \mathcal{O}(\Delta^{2}) 
                              \nonumber\\
 & \equiv & H_{0}(t) + \sum_{j=1}^{3}\mathcal{G}_{j}\Delta F_{j}(t) ,
\label{Hamdef}
\end{eqnarray}
where $\calG_{j} = \left. \delta H/\delta F_{j}\right|_{\bff_{0}}$ is an 
$N\times N$ matrix obtained by taking the functional derivative of 
$H[\bff (t)]$ with respect to $F_{j}(t)$ evaluated at $\bff_{0}(t)$, and 
$N$ is the dimension of the Hilbert space. For example, suppose $H(t)$ is the
Zeeman Hamiltonian $H(t) = -\bldsig\cdot\bff (t)$, where the $1,2,3$ components 
of $\bldsig$ are the $x,y,z$ Pauli matrices, respectively. Then, a simple 
calculation gives $\calG_{j} = -\bldsig_{j}$.

The Schrodinger equation for the propagator $U(t)$ is ($\hbar = 1$)
\begin{equation}
i\frac{dU}{dt} = H(t)U .
\label{propSch}
\end{equation}
For $H(t) = H[\bff (t)]$, the propagator $U(t)$ becomes a functional of the
control field $\bff (t)$. Throughout this paper we assume that the nominal 
control field $\bff_{0}(t)$ acts for a time $T$ and gives rise to a propagator 
$U_{0}(t)$ which provides a good approximation $U_{0}(t=T)$ to a target 
gate $U_{tgt}$ \cite{foot1}. We introduce the gate modification $\delta U(t)$ 
by writing $U(t) = U_{0}(t)\delta U(t)$. Inserting Eq.~(\ref{Hamdef}) into 
Eq.~(\ref{propSch}), and substituting for $U(t)$ gives the equation of 
motion for $\delta U(t)$:
\begin{eqnarray}
i\frac{d}{dt}\delta U & = & 
      \left[ \sum_{j = 1}^{3} \left(  U^{\dagger}_{0} \calG_{j} U_{0}\right) 
             \Delta F_{j} \right]\delta U  + \calO (\Delta^{2}) \nonumber\\
 & = & \left[ \sum_{j=1}^{3} \barG_{j} \Delta F_{j}\right] \delta U .
\label{inteqn}
\end{eqnarray}
Here $\barG_{j} = U^{\dagger}_{0}(t)\calG_{j}U_{0}(t)$ is an $N\times N$
matrix; and the initial condition $\delta U(0) = I$ follows from the definition 
of $\delta U(t)$ and $U(0) = U_{0}(0) = I$. By assumption, $U_{0}(t)$ already 
gives a good approximation to the target gate $U_{tgt}$, and so we look for 
a gate modification $\delta U(t)$ that is close to the identity: $\delta U(t) = 
I -i\delta A(t) + \mathcal{O}(\Delta^{2})$. Note that $\delta A(t)$ is 
Hermitian, and $\delta A(0) = 0$. Substituting this expression for $\delta U(t)$
into Eq.~(\ref{inteqn}) gives 
\begin{equation}
\frac{d}{dt}\delta A =  \sum_{j=1}^{3} \overline{G}_{j}\Delta F_{j}  + 
                                           \mathcal{O}(\Delta^{2}).
\label{inteq2}
\end{equation}
It proves useful to write the $N\times N$ matrix $\delta A(t)$ as an
$N^{2}$-component column vector $\Delta\bldx (t)$. This is done by concatenating
the columns $\{\delta A_{\:\mbox{\boldmath $\cdot$} , j}(t) : j = 1, \cdots , 
N\}$ of $\delta A(t)$ into a 
single column vector: 
\begin{equation}
\Delta\bldx (t)  =  \left( \begin{array}{c}
                                      \delta A_{\:\bldcdot , 1}(t)\\
                                       \vdots \\
                                        \delta A_{\:\bldcdot , N}(t)
                                 \end{array} \right) . \label{defs1/2}
\end{equation}
We also construct an $N^{2}\times 3$ matrix $G(t)$ as follows. First we take 
each $N\times N$ matrix $\barG_{j}(t)$ and convert it into an $N^{2}$-component 
column vector $\bfG_{j}(t)$ as described above. We then insert $\bfG_{j}(t)$ 
into the $j$-th column of $G(t)$:
\begin{eqnarray}
G(t)  & = & \left(   \begin{array}{ccc}
                        \vdots & \vdots & \vdots \\
                        \bfG_{1}(t) & \bfG_{2}(t) & \bfG_{3}(t)\\
                          \vdots & \vdots & \vdots 
                      \end{array}
           \right).
\label{Gdef}
\end{eqnarray}
Finally, we introduce the column vector $\Delta\bff (t)$: 
\begin{equation}
\Delta\bff (t) = \left(  \begin{array}{c}
                                       \Delta F_{1}(t)\\
                                       \Delta F_{2}(t)\\
                                       \Delta F_{3}(t)
                                   \end{array}
                         \right).
\label{Fdef}
\end{equation}
With these definitions, Eqn.~(\ref{inteq2}) is transformed into the equation of 
motion for $\Delta \bldx (t)$: 
\begin{eqnarray}
\frac{d}{dt}\Delta\bldx  & = & G(t) \Delta\bff (t)  ,
\label{schrodinger}
\end{eqnarray}
where the rhs is the matrix product of Eqs.~(\ref{Gdef}) and (\ref{Fdef}),
and the initial condition $\Delta\bldx (0) = 0$ follows from $\delta A(0) = 0$.

\subsection{Dynamical optimization problem}
\label{sec2b}

In optimal control theory the problem is to determine a control field profile
$\bff_{\ast} (t)$ that optimizes system performance relative to a set of design 
criteria. A cost function is introduced that quantifies the degree to which a 
particular assignment of the control and system variables satisfies these  
criteria, with an optimal assignment being one of minimum cost \cite{foot2}.
The cost function $J$ used in our gate optimization contains three 
contributions: (i)~a terminal cost $J_{1}$ that vanishes when the final 
propagator $U(t=T)$ equals the target gate $U_{tgt}$; (ii)~an integral cost 
$J_{2}$ that insures the control field and state modifications, 
respectively, $\Delta\bff (t)$ and $\Delta\bldy (t)$ remain small at 
all times; and (iii)~a Lagrange multiplier integral cost $J_{3}$ that insures 
the optimization does not violate the Schrodinger dynamics of $\Delta\bldy 
(t)$.\\

\textbf{1.~Terminal cost $J_{1}$:} As shown in Ref.~\cite{trp5}, and summarized
in Appendix~\ref{appendixTRP},
\begin{displaymath}
Tr\, P = Tr\left[\,\left( U^{\dagger}(T) - U^{\dagger}_{tgt}\right)\left( 
                  U(T) -   U_{tgt}\right)\,\right],
\end{displaymath}
is a convenient upper bound on the gate error probability $P_{e}$ which is 
clearly minimized when $U(T) = U_{tgt}$. We will use it as a terminal cost:
\begin{eqnarray}
J_{1} & = &  Tr\left[\,\left( U^{\dagger}(T) - U^{\dagger}_{tgt}\right)
                                 \left( U(T) -  U_{tgt}\right)\,\right] .
\end{eqnarray}
The cost $J_{1}$ enforces the criterion that $U(T) = U_{tgt}$ softly, allowing 
it to be violated, but penalizing violations with non-zero cost.
By assumption, $U_{0}(T)$ is a good approximation for $U_{tgt}$, and so
$U^{\dagger}_{0}(T)U_{tgt} = I -i\delta\beta + \mathcal{O}(
\Delta^{2})$, where $\delta\beta$ is Hermitian. Recall that $U(t) = 
U_{0}(t)\delta U(t)$ and $\delta U(t) = I -i\delta A(t) +\mathcal{O}(
\Delta^{2})$. Expanding $J_{1}$ to second order gives: 
\begin{equation}
J_{1}  =  Tr\left[\, \left(\delta A^{\dagger}(T)-\delta\beta^{\dagger}\right)
                      \left( \delta A(T)-\delta\beta\right)\,\right] .
\label{J2int}
\end{equation}
If we write $\delta\beta$ as a (constant) $N^{2}$-component vector 
$\Delta\bldb$ as was done with $\delta A(t)$ in Eq.~(\ref{defs1/2}), we can
re-write $J_{1}$ as the product of a row and column vector
\begin{equation}
J_{1} = \left(\Delta\bldx^{\dagger}(T) - \Delta\bldb^{\dagger}\right)
               \left( \Delta\bldx (T) - \Delta\bldb\right) .
\end{equation}
Defining the column vector $\Delta\bldy (t)$ as
\begin{equation}
\Delta\bldy (t) = \Delta\bldx (t) - \Delta\bldb, 
\label{ydef}
\end{equation}
$J_{1}$  becomes the square-magnitude of $\Delta\bldy (T)$
\begin{equation}
J_{1} = \Delta\bldy^{\dagger}(T)\Delta\bldy (T) .
\end{equation}
Note that since $\Delta\bldb$ is a constant vector,
$\Delta\bldy (t)$ also satisfies Eq.~(\ref{schrodinger}):
\begin{equation}
\frac{d}{dt}\Delta\bldy = G \Delta\bff   .
\label{yscheq}
\end{equation}
The initial condition for Eq.~(\ref{yscheq}) is $\Delta\bldy (0) = 
-\Delta\bldb$ which follows from Eq.~(\ref{ydef}) and $\Delta\bldx (0) = 0$. 
It proves convenient in the following to work with $\Delta\bldy (t)$ instead of
$\Delta\bldx (t)$. \\

\textbf{2.~Integral cost $J_{2}$:} The second cost term $J_{2}$ is an integral 
cost that penalizes large values of $\Delta\bff (t)$ and $\Delta\bldy (t)$ for
all times $t$:
\begin{equation}
J_{2} = \int_{0}^{T} dt \left[ \Delta\bldy^{\dagger}(t) Q(t)
               \Delta\bldy (t) + \frac{1}{2}\Delta\bff^{T}(t) R(t)
                \Delta\bff (t)\right] .
\end{equation}
Here $Q(t)$ and $R(t)$ are positive-definite Hermitian matrices, but otherwise, 
are at our disposal \cite{foot3}. The cost $J_{2}$ is minimized by vanishing 
state and control modifications $\Delta\bldy (t)=0$ and $\Delta\bff (t)=0$. 
Non-vanishing $\Delta\bldy (t)$ and $\Delta\bff (t)$ are allowed to occur, but 
they are penalized with non-zero cost. Thus $J_{2}$ acts to softly enforce 
the criterion of small state and control modifications.\\

\textbf{3.~Integral cost $J_{3}$:} Finally, we require that the optimization 
obey the Schrodinger dynamics of $\Delta\bldy (t)$. This criterion is enforced 
as a hard constraint which cannot be violated by introducing a Lagrange 
multiplier $\Delta\bldlam (t)$:
\begin{eqnarray}
J_{3} & = & \int_{0}^{T} dt \left[ \Delta\bldlam^{\dagger}(t) \left\{    
                  G(t)\Delta\bff (t)   - \Delta\dot{\bldy}(t)\right\} 
                               + h.~c.\right] \nonumber \\
 & = & -\left. \Delta\bldlam^{\dagger}\Delta\bldy\right|_{0}^{T} 
                   \nonumber\\
 & &  \hspace{0.1in} + \int_{0}^{T} dt \left[ \left( \Delta\bldlam^{\dagger}(t)
                        G(t)  \Delta\bff (t)  +\Delta\dot{\bldlam}^{\dagger}(t)
                     \Delta\bldy (t)  \right) \right.\nonumber\\ 
 & & \hspace{2.0in} \left. +\:  h.~c.\:\right]. 
\label{J3cost}
\end{eqnarray}
Note that we have done an integration by parts in going from the first to the 
second line; a dot over a symbol indicates a time-derivative; and $h.~c.$ 
indicates the Hermitian conjugate of the preceeding term. \\

\textbf{4.~Total cost $J$:} Combining all three costs gives
\begin{eqnarray}
J & = & \left[ \Delta\bldy^{\dagger}(T)\Delta\bldy (T)  
             - \Delta\bldlam^{\dagger}(T)\Delta\bldy (T)\right] 
                           \nonumber \\
  &  & \hspace{0.1in}+ \int_{0}^{T} dt \left[ 
                \Delta\bldy^{\dagger}(t) Q(t) \Delta\bldy (t) + 
                \frac{1}{2}\Delta\bff^{T}R(t)\Delta\bff (t)\right] \nonumber\\
 & & \hspace{0.2in} + \int_{0}^{T} dt\left[ \left( 
              \Delta\dot{\bldlam}^{\dagger}(t)
              \Delta\bldy (t) + \Delta\bldlam^{\dagger}(t)
                G(t)\Delta\bff (t)\right)  \right.\nonumber\\
 & & \hspace{2.0in}\left.  +\:  h.~c.\: \right].
\end{eqnarray}
As we shall see in Section~\ref{sec2c}, appropriate variation of $J$ gives 
the equations that govern the optimization, including the feedback
control law. Note that we have dropped the $\Delta\bldlam^{\dagger}(0)
\Delta\bldy (0)$ contribution to $J$ that arises from the surface term in 
Eq.~(\ref{J3cost}) as it has zero variation since $\Delta\bldy (0) =
-\Delta\bldb$ is a constant with zero variation. 

\subsection{Euler-Lagrange equations for optimal control}
\label{sec2c}

A necessary condition for optimal control is that the first-order variation 
of the cost function $J$ vanish. This is most easily worked out by taking 
functional derivatives of $J$ with respect to $\Delta\bldy (t)$, $\Delta 
\bff (t)$, and $\Delta\bldlam (t)$, and setting these derivatives equal to 
zero. This leads to the equations of motion that govern the optimization.
It follows automatically from the positive-definite quadratic nature of $J$ 
that its second-order variation is positive, making the extremum solution 
found from the first-order variation the desired minimum cost solution.\\

\textbf{1.~Variation of $\Delta\bldy (t)$:} Taking the functional derivative of 
$J$ with respect to $\Delta\bldy (T)$ and setting the result equal to zero gives
\begin{displaymath}
\Delta\bldy^{\dagger}(T) - \Delta\bldlam^{\dagger}(T) = 0.
\end{displaymath}
Solving for $\Delta\bldlam (T)$ gives:
\begin{equation}
\Delta\bldlam (T) = \Delta\bldy (T).
\label{LEOM1}
\end{equation}
Next, taking the functional derivative of $J$ with respect to $\Delta\bldy (t)$ 
and setting the result equal to zero gives
\begin{displaymath}
\Delta\bldy^{\dagger}(t)Q(t) + \Delta\dot{\bldlam}^{\dagger}(t)  = 0.
\end{displaymath}
Solving for $\Delta\dot{\bldlam}(t)$ gives (recall $Q(t)$ is Hermitian):
\begin{equation}
\frac{d}{dt}\Delta\bldlam (t) = - Q(t)\Delta\bldy (t) .
\label{LEOM2}
\end{equation}
Eqs.~(\ref{LEOM1}) and (\ref{LEOM2}) define an initial value problem for
the Lagrange multiplier $\Delta\bldlam (t)$, where the ``initial'' time is
$t=T$. Note that taking the functional derivative of $J$ with respect to 
$\Delta\bldy^{\dagger}(t)$ simply gives the adjoint of these equations and so
provides no new information.\\

\textbf{2.~Variation of $\Delta\bff (t)$:} Taking the functional derivative of 
$J$ with respect to $\Delta\bff (t)$ and setting it equal to zero gives:
\begin{displaymath}
\Delta\bff^{T}(t) R(t) + \Delta\bldlam^{\dagger} G(t) = 0.
\end{displaymath}
Solving for $\Delta\bff (t)$ gives (recall $R(t)$ is positive-definite and 
Hermitian):
\begin{equation}
\Delta\bff (t) = - R^{-1}(t) G^{\dagger}(t)\Delta\bldlam (t) .
\label{LEOM3}
\end{equation}
Eq.~(\ref{LEOM3}) relates the control modification $\Delta\bff (t)$ to the
Lagrange multiplier $\Delta\bldlam (t)$. Note that for the second strategy 
presented in Section~\ref{sec2d}, this equation will be transformed into a 
feedback control law.\\

\textbf{3.~Variation of $\Delta\bldlam (t)$:} By design, $J_{3}$ was added
to the cost function to insure that the Schrodinger dynamics of $\Delta\bldy 
(t)$ is not violated by the optimization process. Taking the functional 
derivative of the first line of Eq.~(\ref{J3cost}) and setting the result 
equal to zero gives
\begin{equation}
\frac{d}{dt}\Delta\bldy (t)  -G(t)\Delta\bff (t) = 0 ,
\label{LEOM4}
\end{equation}
which is Eq.~(\ref{yscheq}) as required. We have already seen that its initial 
condition is
\begin{equation}
\Delta\bldy (0) = -\Delta\bldb .
\label{LEOM5}
\end{equation}

\subsection{Solution strategies}
\label{sec2d}

Here we describe two strategies for solving the Euler-Lagrange equations of 
motion for optimal control (Eqs.~(\ref{LEOM1})-(\ref{LEOM5})). Each strategy 
provides a way to determine $\Delta\bldlam (t)$ without directly integrating 
Eqs.~(\ref{LEOM1})--(\ref{LEOM2}). The first is based on an ansatz for the 
Lagrange multiplier $\Delta\bldlam (t)$, while the second relates $\Delta
\bldlam (t)$ to $\Delta\bldy (t)$ through the Ricatti matrix $S(t)$.

In Section~\ref{sec3} and Appendix~\ref{appendixRemainingResults} we use
our neighboring optimal control formalism to improve the performance of all 
gates in the universal set of gates introduced in Appendix~\ref{appendixA3}. 
Strategy~1 will be used to improve all one-qubit gates, while Strategy~2 will 
be used to improve the sole two-qubit gate in the set.\\

\textbf{Strategy~1 -- Lagrange multiplier ansatz:} This approach to solving 
the Euler-Lagrange (EL) equations for optimal control is based on the following 
ansatz for the Lagrange multiplier:
\begin{equation}
\Delta\bldlam (t) = -\exp\left[ -(t +t_{0}/2)/10\right] \bldw ,
\label{LMansatz}
\end{equation}
where $-T/2 \leq t\leq T/2$, and $\bldw$ is a constant vector that is determined
by demanding that: (i)~the gate modification $\delta A(t) = i[\delta U(t)  - I]$
satisfies the Schrodinger equation (viz.\ Eq.~(\ref{inteq2})); and (ii)~$\delta 
A(T/2)=\delta\beta + \mathcal{O}(\Delta^{2})$, where $\delta\beta = 
i[U^{\dagger}_{0}(T/2)U_{tgt} - I] + \calO (\Delta^{2})$ (see 
Section~\ref{sec2b}). Note that, because of the second requirement, 
\begin{eqnarray*}
\delta U(T/2) & = & I -i\delta A(T/2) + \calO (\Delta^{2})\\
 & = & I - i\delta\beta + \calO (\Delta^{2})\\
 & = & U^{\dagger}_{0}(T/2)U_{tgt} + \calO(\Delta^{2}) ,
\end{eqnarray*}
and consequently, the new gate $U(T/2) = U_{0}(T/2)\delta U(T/2)$ satisfies:
\begin{eqnarray}
U(T/2) &=& U_{0}(T/2)\left[ U^{\dagger}_{0}(T/2)U_{tgt}
                                       \right]\nonumber\\
 & = & U_{tgt} + \calO (\Delta^{2}) .
\label{secreq}
\end{eqnarray}
Thus, by choosing $\bldw$ in this way, we insures that EL 
Eqs.~(\ref{LEOM4}) and \ref{LEOM5}) are satisfied, and the new gate 
$U(T/2)$ is the target gate $U_{tgt}$ to second-order in small quantities.

We choose $R(t) = I$ so that Eq.~(\ref{LEOM3}) gives the control modification:
\begin{equation}
\Delta\bff (t ) = \exp\left[ -(t +T/2)/10\right] G^{\dagger}(t )
                                         \bldw .
\label{conmod1}
\end{equation}
Once $\bldw$ is determined, EL Eq.~(\ref{LEOM3}) is satisfied.

Finally, choosing $Q(t) $ to be a diagonal matrix, Eq.~(\ref{LEOM2}) 
determines $Q(t)$ from the ansatz for $\Delta\bldlam (t)$ and the solution
$\Delta\bldy (t)$ of Eqs.~(\ref{LEOM4}) and (\ref{LEOM5}). With this choice,
EL Eq.~(\ref{LEOM2}) is satisfied. Thus, once $\bldw$ is known, the 
strategy's construction insures that all EL equations are satisfied, and yields
the control and gate modifications $\Delta\bff (t)$ and $\Delta\bldy (t)$. 
Note that Strategy~1 has the following significant benefit. By introducing an 
ansatz for $\Delta\bldlam (t)$, computation of the control and gate 
modifications $\Delta\bff (t)$ and $\Delta\bldy (t)$ becomes independent of 
$Q(t)$. Thus Strategy~1 does not actually require $Q(t)$ to be computed. 
We now describe how $\bldw$ is determined.

We begin with Eq.~(\ref{inteq2}), together with Eq.~(\ref{conmod1}):
\begin{eqnarray}
\frac{d}{dt}\delta A & = & \sum_{j=1}^{3} \barG_{j}\Delta F_{j}\nonumber\\
  & = & \exp\left[ -(t +T/2)/10\right] \sum_{j=1}^{3}\barG_{j}
                       \left( G^{\dagger}\bldw \right)_{j} .
\label{delAeqn}
\end{eqnarray}
In Appendix~\ref{appendixB} we show that
\begin{equation}
\sum_{j=1}^{3}\barG_{j}\left( G^{\dagger}\bldw \right)_{j} = 
               \left(  \begin{array}{cc}
                            w_{1}-w_{4} & 2w_{3}\\
                            2w_{2} & w_{4} - w_{1}
                        \end{array} \right)  .
\label{bigsurprise}
\end{equation}
Note that in deriving this result we explicitly assume that our quantum
system is a single qubit whose dynamics is driven by the Zeeman Hamiltonian 
$H(t) = -\bldsig \cdot\bff (t)$. Using Eq.~(\ref{bigsurprise}) in 
Eq.~(\ref{delAeqn}) gives
\begin{eqnarray}
\lefteqn{\frac{d}{dt} \left(  \begin{array}{cc}
                                         \delta A_{11} & \delta A_{12} \\
                                        \delta A_{21} & \delta A_{22}
                                   \end{array} \right)  = }\nonumber\\
 & & \hspace{0.1in}  \exp\left[ -(t + T/2)/10\right]
        \left(   \begin{array}{cc}
                       w_{1} - w_{4} & 2w_{3}\\
                       2w_{2} & w_{4} - w_{1}
                   \end{array}  \right) .
\label{inteqnsec21}
\end{eqnarray}
This equation is easily integrated, with the result:
\begin{subequations}
\begin{eqnarray}
\delta A_{11}(t) & = & 10\left( w_{1}-w_{4}\right)\calA (t) \\
\delta A_{21}(t) & = & 20 w_{2}\,\calA (t) \\
\delta A_{12}(t) & = & 20w_{3}\,\calA (t) \\
\delta A_{22}(t) & = & 10\left( w_{4}-w_{1}\right)\calA (t),
\end{eqnarray}
\end{subequations}
where
\begin{equation}
\calA (t) = 1 - \exp\left[ -(t + T/2)/10\right] .
\end{equation}
For the one-qubit gate simulations presented in Section~\ref{sec3} and 
Appendix~\ref{appendixRemainingResults} we have $T = 160$ \cite{foot6}. 
Thus $\calA (T/2) = 1 - \exp (-16) = 1 + \calO (10^{-7})$. Combining this
with the requirement that $\delta A(T/2) = \delta\beta$ gives
\begin{subequations}
\label{allequations}
\begin{eqnarray}
w_{1}-w_{4} & = & \frac{\delta\beta_{11}}{10} \\
w_{2} & = & \frac{\delta\beta_{21}}{20} \\
w_{3} & = & \frac{\delta\beta_{12}}{20} \\
w_{4}-w_{1} & = & \frac{\delta\beta_{22}}{10} .
\end{eqnarray}
\end{subequations}
Recall that $U^{\dagger}_{0}(T/2)U_{tgt} = I -i\delta\beta + 
\calO (\Delta^{2})$ so that
\begin{eqnarray}
\hspace{-0.1in} Tr\left[ U^{\dagger}_{0}(T/2)U_{tgt}\right] & = &
          2 - i\, Tr\,\delta\beta + \calO (\Delta^{2}) \nonumber \\
  & = & 2 -i\left(\delta\beta_{11} +\delta\beta_{22}\right) +
                 \calO (\Delta^{2}) .
\end{eqnarray}
In Appendix~\ref{appendixA3} we show that for all one-qubit gates of interest
in this paper, $Tr\left[ U^{\dagger}_{0}(T/2)U_{tgt}\right] = 2 +
\calO (\Delta^{2})$ so that
\begin{equation}
\delta\beta_{11}+\delta\beta_{22} = 0.
\label{minoreqn}
\end{equation}
Combining Eq.~(\ref{minoreqn}) with the choice $w_{1} = -w_{4}$, 
reduces Eqs.~(\ref{allequations}) to
\begin{equation}
\bldw = \frac{\Delta\bldb}{20} ,
\label{bldwfinalform}
\end{equation}
where, recall,
\begin{eqnarray}
\Delta\bldb & = & \left(  \begin{array}{c}
                                           \delta\beta_{11} \\
                                           \delta\beta_{21} \\
                                           \delta\beta_{12} \\
                                           \delta\beta_{22} \\
                                       \end{array} \right) .
\label{delbetadef2}
\end{eqnarray}
Eqs.~(\ref{bldwfinalform}) and (\ref{delbetadef2}), together with $\delta\beta =
i[U^{\dagger}_{0}(T/2)U_{tgt} - I]$, determine $\bldw$. As was noted above, 
this then determines the control modification $\Delta\bff (t)$, and solution
of the Schrodinger equation determines $\Delta\bldy (t)$ which gives 
the gate modification $\delta U(t)$. The new control field is $\bff (t) = 
\bff_{0}(t)+\Delta\bff (t)$, and the new gate is $U(T/2) = U_{0}(T/2)\delta 
U(T/2)$. We implement Strategy~1 in Section~\ref{sec3} and 
Appendix~\ref{appendixRemainingResults} to improve the one-qubit gates 
in the universal quantum gate set introduced in Appendix~\ref{appendixA3}.
\\

\textbf{Strategy~2 -- Ricatti equation and the control gain matrix:} From 
Eq.~(\ref{LEOM2}) we see that $\Delta\bldy (t)$ acts as the source for the 
Lagrange multiplier $\Delta\bldlam (t)$. We look for a solution of 
Eq.~(\ref{LEOM2}) of the form
\begin{equation}
\Delta\bldlam (t) = S(t)\Delta\bldy (t) ,
\label{riccmatdef}
\end{equation}
where $S(t)$ is known as the Ricatti matrix. Note that once $S(t)$ has been
determined, Eq.~(\ref{LEOM3}) becomes the feedback control law
\begin{eqnarray}
\Delta\bff (t) & = & -R^{-1}(t) G^{\dagger}(t) S(t)\Delta\bldy (t)
                                                     \nonumber\\
 & = & - C(t)\Delta\bldy (t) 
\label{feedbacklaw}
\end{eqnarray}
which relates the state modification $\Delta\bldy (t)$ to the control 
modification $\Delta\bff (t)$. The matrix $C(t) = R^{-1}(t) G^{\dagger}(t) 
S(t)$ is known as the control gain matrix. To obtain the equation of motion 
for $S(t)$ we differentiate Eq.~(\ref{riccmatdef}), and then use 
Eqs.~(\ref{LEOM2}) and (\ref{LEOM4}) to substitute for $\Delta\dot{\bldlam}$ 
and $\Delta\dot{\bldy}$. One finds
\begin{eqnarray}
\dot{S}\Delta\bldy & = & \Delta\dot{\bldlam} - S\Delta\dot{\bldy} 
                                                 \nonumber\\
 & = & -  Q\Delta\bldy 
 -SG\Delta\bff  \nonumber \\
 & = & -Q\Delta\bldy - S G(-R^{-1}G^{\dagger}S\Delta\bldy ) \nonumber\\
 & = & \left[ -Q + SGR^{-1}G^{\dagger}S\right]\, \Delta\bldy .
\label{almostricc}
\end{eqnarray}
Identifying the coefficients of $\Delta\bldy$ on both sides of 
Eq.~(\ref{almostricc}) gives the Ricatti equation
\begin{equation}
\frac{dS}{dt} = -Q + SGR^{-1}G^{\dagger}S .
\label{ricceq}
\end{equation}
The ``initial'' condition for $S(T)$ is found from Eqs.~(\ref{LEOM1}) and
(\ref{riccmatdef}):
\begin{displaymath}
\Delta\bldy (T) = S(T)\Delta\bldy (T),
\end{displaymath}
from which it follows that
\begin{equation}
S(T) = I.
\label{riccIC}
\end{equation}

 Note that by introducing the Ricatti matrix $S(t)$ we have transformed the
problem of finding the Lagrange multiplier $\Delta\bldlam (t)$ to that of 
finding $S(t)$. This is a good strategy as the Ricatti equation is independent 
of both $\Delta\bldy (t)$ and $\Delta\bff (t)$ and so can be solved once and for
all. This is not the case with Eq.~(\ref{LEOM2}). The equations that determine 
the path and control modifications $\Delta\bldy (t)$ and $\Delta\bff (t)$ 
are thus Eqs.~(\ref{ydef}), (\ref{LEOM4}), (\ref{LEOM5}), (\ref{feedbacklaw}), 
(\ref{ricceq}), and (\ref{riccIC}). Note that by substituting the feedback 
control law (Eq.~(\ref{feedbacklaw})) into Eq.~(\ref{LEOM4}) we obtain
\begin{equation}
\frac{d}{dt}\Delta\bldy = - GC\Delta\bldy .
\label{newLEOM4}
\end{equation}
Once the Ricatti matrix $S(t)$ is known, the control gain matrix $C(t)$ is known,
and Eq.~(\ref{newLEOM4}) can then be integrated for $\Delta\bldy (t)$. With 
$\Delta\bldy (t)$ in hand, Eq.~(\ref{feedbacklaw}) determines the control 
modification $\Delta\bff (t)$, and so the improved control 
$\bff (t) = \bff_{0}(t) + \Delta\bff (t)$. Note that if all the 
eigenvalues of $GC$ are positive, then $\Delta\bldy (t\rightarrow\infty ) = 0$,
and so from Eq.~(\ref{ydef}), that $\Delta\bldx (t\rightarrow\infty ) = 
\Delta\bldb$. This, in turn implies that $\delta U(t\rightarrow\infty ) = 
U^{\dagger}_{0}U_{tgt}$, and finally, $U(t\rightarrow\infty ) = U_{tgt}$ as 
desired.

\section{Example: Improving a universal quantum gate set}
\label{sec3}

Having constructed in Section~\ref{sec2} a general theoretical framework for 
improving the performance of a good quantum gate, we now illustrate its use by 
applying it to a universal set of quantum gates that has appeared in the literature 
\cite{trp1,trp2,trp3,trp4,trp5,trp6,trp7,trp8}. These gates are implemented 
using a form of non-adiabatic rapid passage known as twisted rapid passage 
(TRP). We stress that the method introduced in Section~\ref{sec2} is not limited
to this particular family of input gates - any other good gate, or set of gates,
could serve as the input for the method. As noted earlier, in the interests of 
clarity, we focus on the Hadamard gate in this Section, and present our 
results for the remaining quantum gates in this set in 
Appendix~\ref{appendixRemainingResults}.

\subsection{Twisted Rapid Passage}
\label{sec3a}

In an effort to make this paper more self-contained, we briefly review the
needed background material on twisted rapid passage (TRP). For a more detailed 
presentation, the reader is directed to Refs.~\cite{trp1,trp5,trp6,trp7}, as 
well as Appendix~\ref{appendixTRP} below.

\subsubsection{TRP and Controllable Quantum Interference} 
\label{sec3a1} 

To introduce TRP \cite{trp1,trp5}, we consider a single-qubit 
interacting with an external control-field $\bff (t)$ via the Zeeman
interaction $H_{z}(t) = -\bldsig\cdot\bff (t)$, where $\sigma_{i}$ are 
the Pauli matrices ($i=x,y,z$). TRP is a generalization of adiabatic rapid 
passage (ARP) \cite{abra}. In ARP, the control-field $\bff (t)$ is slowly 
inverted over a time $T$ with $\bff (t) =at\,\bfhatz + b\,\bfhatx$. 
In TRP, however, the control-field is allowed to twist in the $x$-$y$ plane 
with time-varying azimuthal angle $\phi (t)$, while simultaneously 
undergoing inversion along the $z$-axis: $\bff_{0} (t) = at\,\bfhatz + 
b\,\cos\phi (t)\,\bfhatx +b\,\sin\phi (t)\,\bfhaty $.
Here $-T/2\leq t\leq T/2$, and throughout this paper, we consider 
TRP with \textit{non-adiabatic\/} inversion. As shown in Ref.~\cite{trp5}, 
the qubit undergoes resonance when 
\begin{equation} 
at -\frac{\hbar}{2}\frac{d\phi}{dt} = 0 . 
\label{rescon}
\end{equation} 
For polynomial twist, the twist profile $\phi (t)$ takes the form 
\begin{equation} 
\phi_{n}(t) = \frac{2}{n}Bt^{n} . 
\label{polytwist} 
\end{equation} 
In this case, Eq.~(\ref{rescon}) has $n-1$ roots, though only real-valued roots 
correspond to resonance. Ref.~\cite{trp1} showed that for $n\geq 3$, the 
qubit undergoes resonance multiple times during a \textit{single\/} TRP sweep: 
(i)~for all $n\geq 3$, when $B>0$; and (ii)~for odd $n\geq 3$, when $B<0$. 
For the remainder of this paper we restrict ourselves to $B>0$, and to 
\textit{quartic\/} twist for which $n=4$ in Eq.~(\ref{polytwist}). During 
quartic twist, the qubit passes through resonance at times $t=0,\pm
\sqrt{a/\hbar B}$ \cite{trp1}. It is thus possible to alter the time separating 
the resonances by varying the TRP sweep parameters $B$ and $a$.

Ref.~\cite{trp1} showed that these multiple resonances have a strong 
influence on the qubit transition probability, allowing transitions to be
strongly enhanced or suppressed through a small variation of the sweep 
parameters. Ref.~\cite{fg2} calculated the qubit transition amplitude to all
orders in the non-adiabatic coupling. The result found there can be 
re-expressed as the following diagrammatic series: 
\setlength{\unitlength}{0.04in}
\begin{eqnarray} 
\lefteqn{T_{-}(t) \hspace{0.1in} = \hspace{0.1in}
                 \begin{picture}(10,4) 
                       \put(10,-1.5){\vector(-1,0){3.25}}
                       \put(5,-1.5){\line(1,0){1.75}} 
                       \put(5,-1.5){\vector(0,1){3.25}}
                       \put(5,1.75){\line(0,1){1.75}} 
                       \put(5,3.5){\vector(-1,0){3.25}}
                       \put(0,3.5){\line(1,0){1.75}} 
                 \end{picture} 
\hspace{0.05in} + \hspace{0.05in}
      \begin{picture}(20,4) 
           \put(20,-1.5){\vector(-1,0){3.25}}
           \put(15,-1.5){\line(1,0){1.75}} 
           \put(15,-1.5){\vector(0,1){3.25}}
           \put(15,1.75){\line(0,1){1.75}} 
           \put(15,3.5){\vector(-1,0){3.25}}
           \put(10,3.5){\line(1,0){1.75}} 
           \put(10,3.5){\vector(0,-1){3.25}}
           \put(10,-1.5){\line(0,1){1.75}} 
           \put(10,-1.5){\vector(-1,0){3.25}}
           \put(5,-1.5){\line(1,0){1.75}} 
           \put(5,-1.5){\vector(0,1){3.25}}
           \put(5,1.75){\line(0,1){1.75}} 
           \put(5,3.5){\vector(-1,0){3.25}}
           \put(0,3.5){\line(1,0){1.75}} 
     \end{picture} } \nonumber\\
  & & \hspace{0.8in} +\hspace{0.05in}
           \begin{picture}(30,7)
              \put(30,-1.5){\vector(-1,0){3.25}}
              \put(25,-1.5){\line(1,0){1.75}}
              \put(25,-1.5){\vector(0,1){3.25}}
              \put(25,1.75){\line(0,1){1.75}}
              \put(25,3.5){\vector(-1,0){3.25}}
              \put(20,3.5){\line(1,0){1.75}}
              \put(20,-1.5){\vector(-1,0){3.25}}
              \put(20,3.5){\vector(0,-1){3.25}}
              \put(20,-1.5){\line(0,1){1.75}}
              \put(15,-1.5){\line(1,0){1.75}}
              \put(15,-1.5){\vector(0,1){3.25}}
              \put(15,1.75){\line(0,1){1.75}}
              \put(15,3.5){\vector(-1,0){3.25}}
              \put(10,3.5){\line(1,0){1.75}}
              \put(10,3.5){\vector(0,-1){3.25}}
              \put(10,-1.5){\line(0,1){1.75}}
              \put(10,-1.5){\vector(-1,0){3.25}}
              \put(5,-1.5){\line(1,0){1.75}}
              \put(5,-1.5){\vector(0,1){3.25}}
              \put(5,1.75){\line(0,1){1.75}}
              \put(5,3.5){\vector(-1,0){3.25}}
              \put(0,3.5){\line(1,0){1.75}}
           \end{picture}
\hspace{0.05in} + \hspace{0.05in} \cdots \hspace{0.05in} . 
\label{diagser} 
\end{eqnarray} 
Lower (upper) lines correspond to propagation in the negative (positive) 
energy-level, and the vertical lines correspond to transitions between the 
two energy-levels. The calculation sums the probability amplitudes for all
interfering alternatives \cite{f&h} that allow the qubit to end up in the 
positive energy-level given that it was initially in the negative energy-level.
As we have seen, varying the TRP sweep parameters varies the time 
separating the resonances. This in turn changes the value of each diagram 
in Eq.~(\ref{diagser}), and thus alters the interference between the 
alternative transition pathways. It is the sensitivity of the individual
alternatives/diagrams to the time separation of the resonances that 
allows TRP to manipulate this quantum interference. Zwanziger et al.\ 
\cite{trp2} observed these interference effects in the transition 
probability using NMR and found excellent quantitative agreement between 
theory and experiment. It is this link between interfering quantum alternatives
and the TRP sweep parameters that we believe underlies the ability of 
TRP to drive high-fidelity non-adiabatic one- and two-qubit gates.

\subsubsection{Universal Quantum Gate Set}
\label{sec3a2}

The universal set of quantum gates $\calGU$ that is of interest here 
consists of the one-qubit Hadamard and NOT gates, together with 
variants of the one-qubit $\pi /8$ and phase gates, and the two-qubit
controlled-phase gate. Operator expressions for these gates are:
(1)~Hadamard: $U_{h}=(1/\sqrt{2})\left(\, \sigma_{z}+
     \sigma_{x}\right)$; (2)~NOT: $U_{not} = \sigma_{x}$;
(3)~Modified $\pi /8$: $V_{\pi /8} = \cos\left( \pi /8\right)\,
\sigma_{x} -\sin\left(\pi /8\right)\,\sigma_{y}$;
(4)~Modified phase: $V_{p} = (1/\sqrt{2})\left(\, \sigma_{x}
 -\sigma_{y}\,\right)$; and (5)~Modified controlled-phase:
$V_{cp} = (1/2)\left[ \left( I^{1}+\sigma_{z}^{1}\right) I^{2}
  - \left(I^{1}-\sigma_{z}^{1}\right)\sigma_{z}^{2}\right]$.
The universality of $\calGU$ was demonstrated in Ref.~\cite{trp6}
by showing that its gates could construct the well-known universal 
set comprised of the Hadamard, phase, $\pi /8$, and CNOT gates.

\subsubsection{Simulation Procedure} 
\label{sec3a3} 

As is well-known, the Schrodinger dynamics is driven by a Hamiltonian $H(t)$ 
that causes a unitary transformation $U(t,t_{0})$ to be applied to an initial 
quantum state $|\psi (t_{0})\rangle$. In this paper, it is assumed that the 
Hamiltonian $H(t)$ contains terms that Zeeman-couple each qubit to the TRP 
control-field $\bff_{0} (t)$. Assigning values to the TRP sweep parameters 
$(a,b,B,T)$ fixes the control-field $\bff_{0} (t)$, and in turn, the actual 
unitary transformation $U_{a} = U(t_{0}+T,t_{0}) $ applied to $|\psi (t_{0})
\rangle$. Ref.~\cite{trp6} used optimization algorithms to find TRP sweep 
parameter values that produced an applied one-qubit (two-qubit) gate $U_{a}$ 
that approximates a desired target gate $U_{tgt}$ sufficiently closely that its
error probability (defined below) satisfies $P_{e}<10^{-4}$ ($10^{-3}$)
\cite{foot4}. In the following, the target 
gate $U_{tgt}$ will be one of the gates in the universal set $\calGU$. Since 
$\calGU$ contains only one- and two-qubit gates, our simulations will only 
involve one- and two-qubit systems.

For the \textit{one-qubit simulations}, the nominal Hamiltonian $H_{0}^{1}(t)$
is the Zeeman Hamiltonian $H_{z}(t)$ introduced in Section~\ref{sec3a1}. 
Ref.~\cite{trp5} (see also Appendix~\ref{appendixTRP}) showed that it can 
be written in the following dimensionless form:
\begin{eqnarray}
H_{0}^{1} (\tau ) & = & (1/\lambda)\,\left\{ -\tau\sigma_{z} 
                                 -\cos\phi_{4}(\tau )\sigma_{x} 
                                   -\sin\phi_{4}(\tau )\sigma_{y}\right\}  
                                                           \nonumber\\
 & = & -\bldsig\cdot\bff_{0}(\tau ),
\label{oneqbtHam}
\end{eqnarray}
where $\bff_{0}(\tau )$ is the dimensionless TRP control field; $\tau = (a/b)t$;
$\lambda = \hbar a/b^{2}$; and for quartic twist, $\phi_{4}(\tau ) = (\eta_{4}/2
\lambda )\tau^{4}$, with $\eta_{4}=\hbar Bb^{2}/a^{3}$. In this Section, we 
show how the neighboring optimal control framework introduced in 
Section~\ref{sec2} is applied to improve the performance of the TRP-generated 
Hadamard gate. As the implementation for the remaining one-qubit TRP gates is 
similar, for reasons of clarity, we defer their discussion to 
Appendix~\ref{appendixRemainingResults}.

For the \textit{two-qubit simulations}, the nominal Hamiltonian $H_{0}^{2}(t)$ 
contains terms that Zeeman-couple each qubit to the TRP control-field
$\bff_{0}(t)$, and an Ising interaction term that couples the two qubits. 
Alternative two-qubit interactions can easily be considered, though all 
simulation results presented in this paper assume an Ising interaction between 
the qubits. To break a resonance-frequency degeneracy $\omega_{12}=
\omega_{34}$ for transitions between, respectively, the ground and 
first-excited states ($E_{1}\leftrightarrow E_{2}$) and the second- and 
third-excited states ($E_{3}\leftrightarrow E_{4}$), the term $c_{4}|E_{4}(t )
\rangle\langle E_{4}(t )|$ was added to $H_{2}(t)$. Combining all of these 
remarks, we arrive at the following (dimensionless) two-qubit Hamiltonian 
(see Ref.~\cite{trp6} or Appendix~\ref{appendixTRP} for further details):
\begin{eqnarray}
\lefteqn{H_{0}^{2}(\tau ) =  \left[ -(d_{1}+d_{2})/2+\tau/\lambda
                                               \right]\sigma_{z}^{1}} \nonumber \\
 & & \hspace{0.5in} +\left[ -d_{2}/2+ \tau/\lambda\right] \sigma_{z}^{2} 
             \nonumber \\
  & &   \hspace{0.75in}  -(d_{3}/\lambda )\left[\cos\phi_{4}\sigma_{x}^{1} +
                      \sin\phi_{4}\sigma_{y}^{1}\right] \nonumber \\
 & &  \hspace{1in}-  (1/\lambda )\left[\cos\phi_{4}\sigma_{x}^{2} 
            +  \sin\phi_{4}  \sigma_{y}^{2}\right]  \nonumber \\
   & &     \hspace{0.55in}        -(\pi d_{4}/2)\sigma_{z}^{1}\sigma_{z}^{2} 
                  \hspace{0.1in} +c_{4}|E_{4}(\tau )\rangle\langle 
                        E_{4}(\tau )| .
\label{twoqbtHam}
\end{eqnarray}
Here: (i)~$b_{i} = \hbar\gamma_{i}B_{rf}/2$, $\omega_{i}=\gamma_{i}
B_{0}$, $\gamma_{i}$ is the coupling constant for qubit $i$, and $i=1,2$; 
(ii)~$\tau = (a/b_{2})t$, $\lambda = \hbar a/b_{2}^{2}$, and $\eta_{4}=
\hbar Bb_{2}^{2}/a^{3}$; and (iii)~$d_{1}=(\omega_{1}-\omega_{2})
b_{2}/a$, $d_{2}=(\Delta /a)b_{2}$, $d_{3} = b_{1}/b_{2}$, and 
$d_{4}=(J/a)b_{2}$, where $\Delta$ is a detuning parameter. In the interests 
of clarity, we present our results for the two-qubit modified controlled phase 
gate in Appendix~\ref{appendixRemainingResults}.

Given an applied gate $U_{a}$, a target gate $U_{tgt}$, and the 
initial state $|\psi\rangle$, it is possible to determine (see Ref.~\cite{trp5} 
or Appendix~\ref{appendixTRP}) the error probability $P_{e}(\psi )$ for the 
TRP final state $|\psi_{a}\rangle = U_{a}|\psi\rangle$, relative to the 
target final state $|\psi_{tgt}\rangle = U_{tgt}|\psi\rangle$. The gate error 
probability $P_{e}$ is defined to be the worst-case value \cite{errprobdef} 
of $P_{e}(\psi )$: $P_{e}\equiv \max_{|\psi\rangle}P_{e}(\psi )$. Introducing 
the positive 
operator $P = \left( U_{a}^{\dagger}-U_{tgt}^{\dagger}\right)\left( U_{a}-
U_{tgt}\right) $, Ref.~\cite{trp5} showed that the error probability $P_{e}$ 
satisfies the upper bound $P_{e}\leq Tr\, P$. Once $U_{a}$ is known, $Tr\, P$ 
is easily evaluated, and so it is a convenient proxy for $P_{e}$ which is harder
to calculate. $Tr\, P$ also has the virtue of being directly related to 
the gate fidelity $\calF_{n} = \left(1/2^{n}\right)\, Re\left[\, Tr\left( 
U_{a}^{\dagger}U_{tgt}\right)\,\right]$ , where $n$ is the number of qubits 
acted on by the gate. It is straightforward to show \cite{trp6} that 
$\calF_{n} = 1 -\left(1/2^{n+1}\right)\, Tr\, P$. The simulations calculate 
$Tr\, P$, which is then used to upper bound the gate error probability 
$P_{e}$. Note that minimizing $Tr\, P$ is equivalent to maximizing the gate 
fidelity $\calF_{n}$.

The procedure for solving the EL equations for optimal control was briefly 
described in Section~\ref{sec2d}. The one-qubit TRP gates presented in 
Ref.~\cite{trp8} and the two-qubit TRP gate presented in Ref.~\cite{trp6} 
will serve as the good gates that are to be improved. For the reader's 
convenience, the TRP sweep parameters for these gates are presented in 
Appendix~\ref{appendixA3}, along with their associated gate error probabilities 
and fidelities. For a particular target gate $U_{tgt}$ belonging to $\calGU$ 
(see Section~\ref{sec3a2}), the TRP sweep parameters corresponding to $U_{tgt}$
determine the TRP control field $\bff_{0}(\tau )$ which then 
drives the nominal Hamiltonian $H_{0}(\tau )$ (see Eqs.~(\ref{oneqbtHam}) 
and (\ref{twoqbtHam}) for one- and two-qubit gates, respectively). The nominal 
Hamiltonian in turn produces the initial good approximate gate 
$U_{0}(\tau_{0}/2,-\tau_{0}/2)$ that is to be improved. 
Here $\tau$ is the dimensionless time introduced above, and $\tau_{0} \equiv 
aT/b$. For each gate in $\calGU$, its TRP approximation $U_{0}(\tau_{0}/2,
-\tau_{0}/2)$ is also reproduced in Appendix~\ref{appendixA3}. For the two 
strategies introduced in Section~\ref{sec2d}, the numerical simulation 
implements the following procedure:
\begin{enumerate}
\item For both Strategies, integrate the Schrodinger equation with the nominal 
Hamiltonian $H_{0}(\tau )$ to obtain $U_{0}(\tau_{0}/2,-\tau_{0}/2)$; calculate 
$\Delta\bldb$. For Strategy~1, also calculate $\bldw$.
\item For both Strategies, calculate $\barG_{j}(\tau ) = U^{\dagger}_{0}(\tau )
\calG_{j}U_{0}(\tau)$, where we have abbreviated $U_{0}(\tau , -\tau_{0}/2)$ 
as $U_{0}(\tau )$, and $\calG_{j}(\tau ) = \delta H/\delta F_{j}|_{\bff_{0}
(\tau)}$; form $G(\tau)$. For Strategy~1, skip Step~3, go to Step~4.
\item For Strategy~2, set $R(\tau ) = I_{3\times 3}$ and $S(\tau )= 
I_{16\times 16}$, where $I_{n\times n}$ is the $n\times n$ identity matrix. 
The Ricatti equation then requires $Q(\tau ) = G(\tau )G^{\dagger}(\tau )$. 
The resulting control gain matrix is $C(\tau ) = G^{\dagger}(\tau )$.
\item 
  \begin{enumerate}
\item For Strategy~1,  use Eq.~(\ref{conmod1}) to determine the control
modification $\Delta\bff (\tau )$. 
\item For Strategy~2, solve Eq.~(\ref{newLEOM4}) with initial condition 
Eq.~(\ref{LEOM5}) for $\Delta\bldy (\tau )$; substitute $\Delta\bldy 
(\tau )$ and $C(t)$ into the feedback control law (Eq.~(\ref{feedbacklaw})) 
to determine $\Delta\bff (\tau )$.
  \end{enumerate}
\item For both Strategies, with the improved control field $\bff (\tau ) = 
\bff_{0}(\tau ) + \Delta\bff (\tau )$, numerically integrate the Schrodinger 
equation to determine the new propagator $U(\tau ,-\tau_{0}/2)$, and the 
improved gate $U(\tau_{0}/2,-\tau_{0}/2)$.
\item For both Strategies, calculate $Tr\, P$ for the new gate. This gives: 
(i)~an upper bound on the new gate error probability $P_{e} \leq Tr\, P$, 
and (ii)~the new gate fidelity $\mathcal{F} = 1 - (1/2^{n+1}) Tr\, P$.
\end{enumerate}

\subsection{Ideal Control}
\label{sec3b}

Here we illustrate the use of neighboring optimal control to improve the
performance of a good quantum gate. To avoid obscuring the presentation by 
showing results for all gates in $\calGU$, we instead focus in the remainder of
this section on the one-qubit Hadamard gate. The results for the remaining 
gates in $\calGU$ appear in Appendix~\ref{appendixRemainingResults}. In 
this subsection we examine performance improvements under ideal control, 
while Section~\ref{sec3c} considers the robustness of these improvements 
to some important control imperfections.

\subsubsection{Performance improvement}
\label{sec3b1}

As noted in Section~\ref{sec2d}, we use: (i)~Strategy~1 to determine the 
performance improvements for the one-qubit gates in $\calGU$; and 
(ii)~Strategy~2 for the two-qubit controlled-phase gate. We saw there that 
Strategy~1 produces a one-qubit gate satisfying $U(\tau_{0}/2) = U_{tgt} 
+\calO (\Delta^{2})$. Here we use the numerical simulation procedure described 
in Section~\ref{sec3a3} to determine the small residual error in a one-qubit gate
$U(\tau_{0}/2)$. A comparable discussion for the two-qubit modified 
controlled-phase gate appears in Appendix~\ref{appendixRemainingResults}. 
Thus, for a given one-qubit TRP gate, we use the corresponding 
values of $\lambda$ and $\eta_{4}$ appearing in Table~\ref{tableapp1}
to determine the nominal control field $\bff_{0}(\tau )$. This determines the 
nominal Hamiltonian $H_{0}(\tau ) = -\bldsig\cdot\bff_{0}(\tau )$, and numerical
integration of the Schrodinger equation (see Eq.~(\ref{propSch})) determines 
the nominal state trajectory $U_{0}(\tau )$. Following the simulation protocol, 
$U_{0}(\tau )$ is used to determine $\delta\beta$ and $\bldw$, as well as the 
matrix $G(\tau )$. Eq.~(\ref{conmod1}) is then used to determine the control 
modification $\Delta\bff (\tau )$, and thus the improved control field $\bff 
(\tau ) = \bff_{0}(\tau ) + \Delta\bff (\tau )$. The new Hamiltonian is 
$H(\tau ) = -\bldsig\cdot\bff (\tau )$, and numerical integration of the 
Schrodinger equation determines the improved state trajectory $U(\tau )$. 
The improved one-qubit gate is then $U(\tau_{0}/2)$. With the new gate in 
hand we determine $Tr\, P$ which then provides an upper bound on the gate error 
probability $P_{e}\leq Tr\, P$. If so desired, one can also calculate the gate 
fidelity $\calF = 1 - (1/4) Tr\, P$. 

As noted earlier, we focus our remarks in the remainder of this Section 
on the Hadamard gate. A comparable discussion of the other gates in $\calGU$ 
appears in Appendix~\ref{appendixRemainingResults}. Implementing the above 
numerical simulation protocol using the TRP approximation to the Hadamard gate 
as the starting point returns an improved gate with $Tr\, P = 1.04\times 
10^{-8}$, and thus a gate error probability satisfying $P_{e}\leq 1.04\times 
10^{-8}$. We see that use of neighboring optimal control has produced a 
\textit{four order-of-magntiude\/} reduction in the gate error probability 
compared to the starting TRP gate for which $P_{e}\leq 1.12\times 10^{-4}$. 
The error probability for the improved gate is also 
\textit{four orders-of-magnitude\/} 
less than the target accuracy threshold of $10^{-4}$. Because $P_{e}$ is so 
small, we do not write out the unitary matrix produced by the numerical 
simulation as it agrees with the target Hadamard unitary matrix to $6$ 
significant figures. For completeness, Table~\ref{table1}
\begin{table*}[!htbp]
\centering
 \caption{\label{table1} Simulation results for all target gates in the 
universal set $\calGU$ for ideal control. The first column lists the target 
quantum gates, while the second column lists the $Tr\, P$ upper bound for the 
gate error probability $P_{e}$ for gates whose performance is improved using 
neighboring optimal control (NOC). The third column lists the $Tr\, P$ upper 
bound for the starting TRP gates which do not use NOC. We see that NOC 
has reduced the error probability for all one-qubit gates by 
\textit{four orders-of-magnitude\/}, and by \textit{two orders-of-magnitude\/} 
for the two-qubit controlled-phase gate. The robustness of these reductions 
to control imperfections is examined in Section~\ref{sec3c}. Although not 
included in the Table, the gate fidelity $\calF_{n}$ for an $n$-qubit gate can 
be determined from $Tr\, P$ using $\calF_{n} = 1 -(1/2^{n+1})Tr\, P$.\\}
  \begin{ruledtabular}
  \begin{tabular}{rcc}
  Target Gate\vspace{0.05in} & $P_{e}\leq TrP$ \hspace{0.1in}(with NOC) & 
       $P_{e}\leq TrP$ \hspace{0.1in} (without NOC)\\  \hline
  NOT  &             $\leq 8.58\times 10^{-9}$ & $\leq 6.27\times 10^{-5}$ \\
Hadamard & $\leq 1.04\times 10^{-8}$ & $\leq 1.12\times 10^{-4}$ \\
  Modified $\pi/8$ &  $\leq 1.06\times 10^{-8}$ & $\leq 2.13\times 10^{-4}$\\
  Modified phase  &  $\leq 1.08\times 10^{-8}$  & $\leq 4.62\times 10^{-4}$ \\
Modified controlled-phase & $\leq 5.21\times 10^{-5}$ & $\leq 1.27\times 
                                                                                                  10^{-3}$\\
  \end{tabular}
  \end{ruledtabular}
\end{table*}
gives the $Tr\, P$ upper bound on the gate error probability $P_{e}$ for all 
gates  in $\calGU$, with and without the neighboring optimal control 
improvements. We see that neighboring optimal control reduces the gate error 
probability by: (i)~four orders-of-magnitude for all one-qubit gates in 
$\calGU$; and (ii)~two orders-of-magnitude for the two-qubit modified 
controlled-phase gate. We examine the robustness of these performance gains 
to some important control imperfections in Section~\ref{sec3c}. Before moving 
on to that discussion, we examine in the following subsection, the amount of 
bandwidth needed to realize the control modification $\bff (\tau )$.

\subsubsection{Control field bandwidth}
\label{sec3b2}

We now examine the bandwidth required to realize the control modifications
$\Delta\bff (t)$. We explicitly consider the Hadamard gate in this subsection;
a similar analysis for the remaining target gates in $\calGU$ appears in 
Appendix~\ref{appendixRemainingResults}. To provide context for our
results, we note that arbitrary waveform generators (AWG) are commercially
available with bandwidths as large as $5$ GHz \cite{agilent}.

\begin{figure}[!htb]
\centering
\includegraphics[trim=1.5cm 6.5cm 0 5cm,clip, 
   width=.5\textwidth]{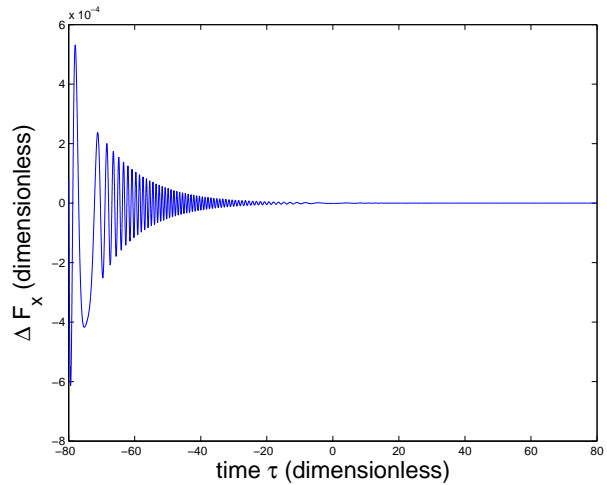}\\
\caption{(Color online) The control modification $x$-component $\Delta F_{x}
(\tau )$ used to
implement a neighboring optimal control improved approximation to the  
Hadamard gate. Here $\tau$ is dimensionless time.\label{fig1}}
\end{figure}

\begin{figure}[!htb]
\centering
\includegraphics[trim=1.5cm 6.5cm 0 5cm,clip,
width=.5\textwidth]{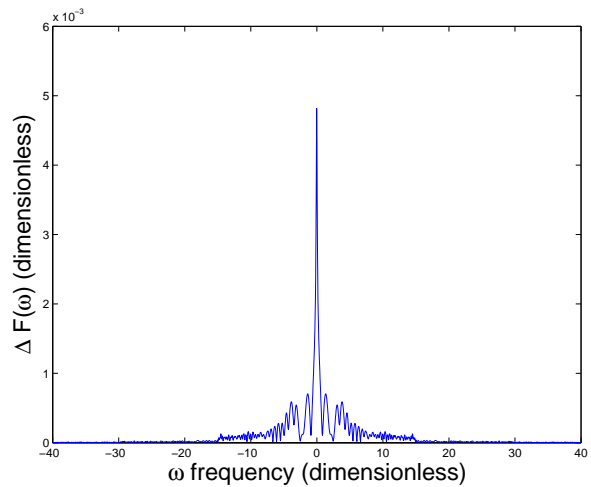}\\
\caption{(Color online) The Fourier transform $\Delta\calF_{x}(\omega )$ of 
$\Delta F_{x} (\tau )$ for the Hadamard Gate as target. Here $\omega$ is
dimensionless frequency.\label{fig2}}
\end{figure}

For a one-qubit target gate, the control modification $\Delta\bff (t)$ is given
by Eq.~(\ref{conmod1}), with $G(t)$ and $\bldw$ determined by the numerical
simulation protocol described in Sections~\ref{sec3a3} and \ref{sec3b1}. 
Figure~\ref{fig1}
shows the $x$-component of the control field modification $\Delta F_{x}(\tau )$ 
as a function of the  dimensionless time $\tau$ for the Hadamard gate as target.
Figure~\ref{fig2}
shows its Fourier transform $\Delta\calF_{x}(\omega )$. We estimate the
(dimensionless) bandwidth of $\Delta\calF_{x}(\omega )$ by determining the
frequency $\omega_{0.1}$ at which $\Delta\calF_{x}(\omega_{0.1})$ is $10\%$ of 
the peak value $\Delta\calF_{x}(0)$. Examination of the numerical data used to 
produce Figure~\ref{fig2} gives $\omega_{0.1} = 4.0$. To convert this into a 
dimensionful bandwidth we suppose that the inversion time $T=1\,\mu \mathrm{s}$.
This corresponds to a dimensionless inversion time of $\tau_{0}=160$ for the
one-qubit gates so that the dimensionful bandwidth $\barom_{0.1}$ is related to 
the dimensionless bandwidth $\omega_{0.1}$ by:
\begin{equation}
\frac{\barom_{0.1}}{\omega_{0.1}}  =  \frac{160}{1\mu\mathrm{s}}\\
     =  160 \,\mathrm{MHz} .
\label{onequbitconvsn}
\end{equation}
Thus the bandwidth needed to implement the control modification $\Delta\bff (t)$
for the Hadamard gate is $\barom_{0.1} = (160\mathrm{MHz})(4.0) = 640
\mathrm{MHz}$. This is well within the range of commercially available AWGs. 
Table~\ref{table2}
\begin{table*}[!htbp]
\centering
 \caption{\label{table2} Bandwidth requirements for neighboring optimal control
improved quantum gates. The dimensionful values assume a one-qubit (two-qubit)
gate time of $1\mu\mathrm{s}$ ($5\mu\mathrm{s}$). Note that the 
bandwidth for the nominal TRP control field $\bff_{0}(t)$ is less than $1\%$ of 
the bandwidth of the control modification $\Delta\bff (t)$. We thus use the 
bandwidth for $\Delta\bff (t)$ as the total bandwidth. Column~1 lists the 
target gate; column~2 the dimensionless bandwidth $\omega_{0.1}$; while
column~3 gives the dimensionful bandwidth $\barom_{0.1}$.\\}
  \begin{ruledtabular}
  \begin{tabular}{rll}
  Target Gate\vspace{0.05in} & $\omega_{0.1}$ \hspace{0.02in}(dimensionless) & 
       $\barom_{0.1}$ \hspace{0.02in} (MHz)\\  \hline
NOT  &             $0.80$ & $130$ \\
Modified $\pi /8$ & $1.3$ & $210$ \\
Modified phase &  $1.9$ & $300$\\
Hadamard  &  $4.0$  & $640$ \\
Modified controlled-phase & $34$ & $820$\\
  \end{tabular}
  \end{ruledtabular}
\end{table*}
lists the bandwidth required to implement the control modification for each of 
the target gates in $\calGU$. The analysis for the other one-qubit gates is 
similar to that of the Hadamard gate, while that of the two-qubit
modified controlled-phase gate has only minor differences. The analysis of 
these other gates appears in Appendix~\ref{appendixRemainingResults}. We 
see that the bandwidth required to implement the neighboring optimal control
performance improvements for all gates in $\calGU$ is squarely within the range 
of existing commercially available AWGs. Note that Eq.~(\ref{onequbitconvsn})
indicates that the dimensionful bandwidth $\barom_{0.1}$ scales as $1/T$ in
the inversion time $T$. Thus, if desired, one can always reduce the 
bandwidth of the control modification $\Delta\bff (t)$ by increasing the 
inversion time (viz.~gate time) $T$.

\subsection{Robustness to control imperfections}
\label{sec3c}

In this subsection we examine the robustness of the neighboring optimal
control (NOC) performance gains found in Section~\ref{sec3b1} to two important
control imperfections. In the interests of clarity, we again focus on the 
Hadamard gate here, and present a similar analysis for the other gates in 
$\calGU$ in Appendix~\ref{appendixRemainingResults}. In Section~\ref{sec3c1} 
we examine the impact of control parameters with finite precision; while in 
Section~\ref{sec3c2} we consider phase noise in the nominal control field.

\subsubsection{Finite-precision control parameters}
\label{sec3c1}

The NOC formalism introduced in this paper requires an input state trajectory
$U_{0}(\tau )$ that yields a good approximation to a target gate $U_{tgt}$. The
control modification $\Delta\bff (\tau )$ determined by the formalism is optimum
for $U_{0}(\tau )$, or equivalently, for the nominal control $\bff_{0}(\tau )$. 
Alteration of the nominal control field $\bff_{0}(\tau )\rightarrow 
\bff_{0}^{\prime}(\tau )$ alters the state trajectory $U_{0}(\tau )\rightarrow 
U_{0}^{\prime}(\tau )$, with the result that the control modification $\Delta
\bff (\tau )$ may no longer be optimal for the altered trajectory 
$U^{\prime}_{0}(\tau )$. Because the hardware used to produce $\bff_{0}(\tau )$
has limited precision, it becomes important to determine the degree of precision
to which the control parameters must be specified if the NOC performance gains 
are to survive the limitation of finite-precision control. 

For the Hadamard gate, Table~\ref{tableapp1} in 
Appendix~\ref{appendixRemainingResults} gives $\lambda = 7.820$ and
$\eta_{4} = 1.792\times 10^{-4}$ as the TRP control parameters that produce
a nominal control field $\bff_{0}(\tau )$, and state trajectory $U_{0}(\tau )$,
for which the gate error probability satisfies $P_{e}\leq 1.12\times 10^{-4}$. 
For these control parameter values, NOC determines the control modification 
$\Delta\bff (\tau )$ (see Section~\ref{sec3b1}) which yields a new gate with 
$P_{e}\leq 1.04\times 10^{-8}$. To examine the robustness of this performance 
improvement, we shift $\lambda$ ($\eta_{4}$) away from its optimum value by 
$1$ in its fourth significant digit, while keeping $\eta_{4}$ ($\lambda$) at  
optimum. This shift causes $\bff_{0}(\tau )\rightarrow
\bff_{0}^{\prime}(\tau )$. We then numerically simulate the Schrodinger 
dynamics driven by the Hamiltonian $H(\tau ) = -\bldsig\cdot\bff^{\prime}
(\tau )$, where the new control field $\bff^{\prime}(\tau ) = 
\bff_{0}^{\prime}(\tau ) + \Delta\bff (\tau )$, and $\Delta\bff (\tau )$ is the 
NOC modification that corresponds to the nominal control field $\bff_{0}
(\tau )$. Tables~\ref{table3}  
\begin{table}[!htbp]
\centering
 \caption{\label{table3}Sensitivity of $TrP$ to a small variation of $\lambda$ 
away from its optimum value for the one-qubit Hadamard gate. For all $\lambda$ 
values, $\eta_{4}$ is maintained at its optimum value $\eta_{4} = 1.792\times 
10^{-4}$. Column~2 (3) shows the variation of $Tr\, P$ when the control 
field includes (omits) the NOC control modification $\Delta\bff(\tau )$.\\}
 \begin{ruledtabular}
\begin{tabular}{ccc}
$\lambda$ & \hspace{0.22in} $TrP$ (with NOC) & \hspace{0.22in}
$TrP$ (without NOC)  \\ \hline
  $7.819$ & $2.62\times 10^{-4}$ & $8.15\times 10^{-4}$ \\
  $7.820$ & $1.04\times 10^{-8}$ & $1.12\times 10^{-4}$ \\
  $7.821$ & $4.44\times 10^{-4}$ & $2.07\times 10^{-3}$ \\
\end{tabular}
\end{ruledtabular}
\end{table}
(\ref{table4})
\begin{table}[!htbp]
\centering
 \caption{\label{table4}Sensitivity of $TrP$ to a small variation of $\eta_4$ 
away from its optimum value for the one-qubit Hadamard gate. For all 
$\eta_{4}$ values, $\lambda$ is maintained at its optimum value $\lambda = 
7.820$. Column~2 (3) shows the variation of $Tr\, P$ when the control 
field includes (omits) the NOC control modification $\Delta\bff(\tau )$.\\}
  \begin{ruledtabular}
\begin{tabular}{ccc}
$\eta_4$ & \hspace{0.22in}$TrP$ (with NOC) & \hspace{0.22in}$TrP$ (without 
NOC)   \\   \hline
 $1.791\times 10^{-4}$ & $5.75\times 10^{-3}$ & $2.86\times 10^{-2}$ \\
 $1.792\times 10^{-4}$ & $1.04\times 10^{-8}$ & $1.12\times 10^{-4}$ \\
  $1.793\times 10^{-4}$ & $7.76\times 10^{-3}$ & $3.11\times 10^{-2}$ \\
\end{tabular}
\end{ruledtabular}
\end{table}
show how the $Tr\, P$ upper bound for the gate error probability $P_{e}$ 
changes due to a small shift in $\lambda$ ($\eta_{4}$) away from its optimum 
value. For comparison, we also show how $Tr\, P$ changes when 
the new control field does not contain the NOC  modification: $\bff^{\prime}
(\tau ) = \bff_{0}^{\prime}(\tau )$. It is clear from these Tables that both 
$\lambda$ and $\eta_{4}$ must be controllable to better than one part in 
$10,000$ if the NOC performance gains are to  be realized. Such control 
parameter precision is attainable using an AWG with $14$-bit vertical resolution
(viz.~one part in $2^{14} = 16,384$). Such AWGs are available commercially
\cite{14bit}. Note that $13$-bit precision corresponds to a precision of one 
part in $2^{13} = 8192$, and so to an uncertainty in the fourth significant 
digit. Thus with less than $14$-bits of precision, Tables~\ref{table3} and
\ref{table4} indicate that the NOC performance gains will be washed out by 
the uncertainty in the least significant digit of $\lambda$ and $\eta_{4}$. 
Lastly, notice that the NOC improved Hadamard gate outperforms the unimproved 
nominal TRP gate, even in the presence of finite precision control parameters. 
This is true for the other gates in $\calGU$ as well.

\subsubsection{Phase/timing jitter}
\label{sec3c2}

Phase jitter arises from timing errors in the clock used by an AWG to produce 
a desired control signal. Ideally, the clock outputs a sequence of ``ticks'' 
with constant time separation $T_{clock}$, derived from an oscillation with 
phase $\phi (t) = 2\pi f_{clock} t$ and frequency $f_{clock} = 1/T_{clock}$. 
A real clock only approximates this ideal behavior. In actuality, the time $T$ 
between ticks is a stochastic process $T = T_{clock} + \delta t$, where the
stochastic timing error $\delta t$ has: (i)~vanishing time-average $\bardelt 
= 0$; and (ii)~a standard deviation $\sigma_{t}=\sqrt{\bardelttwo}$ which 
quantifies the spread of the tick intervals about $T_{clock}$. The spread 
$\sigma_{t}$ is known as timing jitter. The timing error $\delta t$ gives rise 
to a phase error $\delta\phi = (2\pi f_{clock})\delta t$ which has: (i)~zero 
time-average $\bardelphi = 0$; and (ii)~standard deviation $\sigma_{\phi} 
=\sqrt{\bardelphitwo}$ which characterizes the spread about $2\pi$ of the 
phase accumulated between ticks: $\phi = 2\pi f_{clock} T$. The spread
$\sigma_{\phi}$ is known as phase jitter. As $\sigma_{\phi}$ and
$\sigma_{t}$ are two ways of describing the clock timing error, the ratio of 
spread to period for the phase ($\sigma_{\phi}/2\pi$) and the time 
($\sigma_{t}/T_{clock}$) are the same. Equating them, and solving for
$\sigma_{t}$ gives
\begin{equation}
\sigma_{t} = \frac{\sigma_{\phi}}{2\pi f_{clock}} .
\label{jitterconv}
\end{equation}
This expression can be thought of as a change in units from jitter in
radians (viz.~$\sigma_{\phi}$) to jitter in seconds (viz.~$\sigma_{t}$).

Phase jitter is anticipated to affect the performance of the TRP gates used
in our illustration of the NOC formalism. We saw in Section~\ref{sec3a1} 
that the performance of these gates relies on quantum interference effects
that arise during a TRP sweep. In the presence of phase jitter, the TRP twist 
profile $\phi_{4}(\tau ) = (\eta_{4}/2\lambda )\tau^{4}$ develops phase 
noise $\delta\phi (\tau )$ due to the timing error $\delta\tau$ in $\tau$. For 
sufficiently strong phase jitter, this phase noise is expected to wash out the 
interference effects that underlie the good performance of the TRP gates. 
Specifically, since this noise adds to the TRP twist phase $\phi_{4}(\tau ) 
\rightarrow \phi_{4}^{\prime}(\tau ) = \phi_{4}(\tau ) + \delta\phi (\tau )$, 
it causes the (dimensionless) TRP control field $\bff^{\prime}_{0}(\tau ) = 
(1/\lambda )\left[\cos\phi_{4}^{\prime}(\tau )\bfhatx +
\sin\phi_{4}^{\prime}(\tau )\bfhaty +\tau \bfhatz\right]$ to twist incorrectly. 
The control field with the NOC modification is now $\bff^{\prime}(\tau ) = 
\bff_{0}^{\prime}(\tau ) + \Delta\bff (\tau )$, where $\Delta\bff (\tau )$ is 
the neighboring optimal control modification determined for the
 TRP control $\bff_{0}(\tau )$ with \textit{jitter-free\/} twist phase 
$\phi_{4}(\tau )$. It is important to appreciate that the phase noise 
$\delta\phi (\tau )$ is unpredictable and so it is not realistic to assume that 
we can recalculate the control modification $\Delta\bff (\tau )$ so that it is 
optimal for $\bff^{\prime}_{0}(\tau )$ since $\bff^{\prime}_{0}(\tau )$ is not
known until the gate is applied. Thus, for a given target gate, one can only 
calculate the control modification $\Delta\bff (\tau )$ which is optimal for 
the jitter-free TRP control $\bff_{0}(\tau )$, and add it to the noisy TRP 
control $\bff_{0}^{\prime}(\tau )$. Since $\Delta\bff (\tau )$ is not optimal 
for $\bff^{\prime}(\tau )$, the NOC performance improvements are expected 
to be reduced by phase jitter. 

\begin{figure}[!htb]
\centering
\includegraphics[trim=1.5cm 7cm 0 6cm,clip,
width=.5\textwidth]{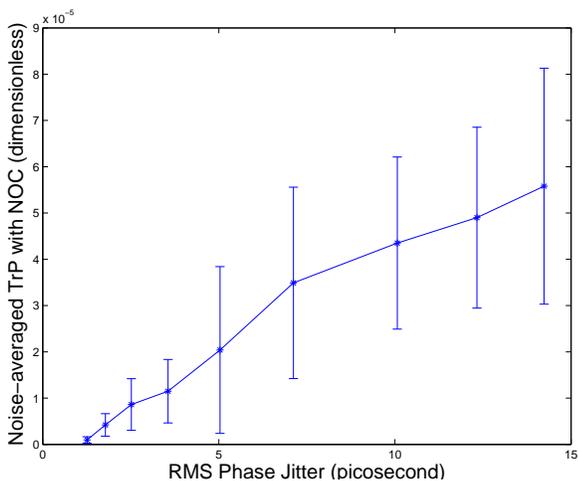}\\
\caption{\label{fig3}(Color online) The noise-averaged value of TrP with 
NOC versus timing jitter $\sigma_{t}=\sigma_{\phi}/(2\pi f_{clock})$ for 
the Hadamard gate. For each $\sigma_{t}$, ten realizations of phase noise 
were generated, and for each realization, gate performance was determined 
by numerical simulation of the Schrodinger dynamics generated by the control 
field $\bff^{\prime}(\tau )$ that includes the noisy TRP nominal control 
$\bff_{0}^{\prime}(\tau )$ and the NOC modification $\Delta\bff (\tau )$ 
(see text). The average and standard deviation were determined for the 
resulting ten $Tr\, P$ values. For each value of $\sigma_{t}$, the average 
of $Tr\, P$ is plotted, and the standard deviation is used to specify the 
error bar. To obtain $\sigma_{t}$, we have assumed that $f_{clock} = 
1\mathrm{GHz}$ (see text).}
\end{figure}

To quantitatively study the effects of phase/timing jitter on the NOC 
performance gains, we modelled the phase noise $\delta\phi (\tau)$ as shot 
noise and used the model to generate numerical realizations of the phase noise 
$\delta\phi (\tau )$.
The details of the model and the protocol used to generate noise realizations 
is described in Appendix~\ref{appendixNoiseModel}. For each noise realization, 
we determined the state trajectory $U(\tau )$ by numerically simulating the 
Schrodinger dynamics generated by the noisy control field $\bff^{\prime}
(\tau )$, and used it to determine the $Tr\, P$ upper bound for the gate error 
probability $P_{e}$. For each target gate $U_{tgt}$ and given value of phase 
jitter $\sigma_{\phi}$ (equivalently, mean phase noise power $\meanP$, see
below), we generated ten realizations of phase noise 
$\delta\phi(\tau )$, and determined the ten corresponding values of $Tr\, P$. 
The average $\langle Tr\, P\rangle$ and standard deviation $\sigma (Tr\, P)$ for
these values was calculated and used to approximate the noise-averaged NOC 
gate performance: $P_{e}\leq\langle Tr\, P\rangle\pm\sigma (Tr\, P)$. We 
carried out simulations for various values of $\sigma_{\phi}$, and present 
our results for the Hadamard gate in Figure~\ref{fig3}.  

To put Figure~\ref{fig3} into context, we note that AWGs with timing jitter
$\sigma_{t} = 5\mathrm{ps}$ and clock frequency $f_{clock} = 1\mathrm{GHz}$ are
commercially available \cite{jitter}. In Appendix~\ref{appendixNoiseModel}
we show that the phase noise variance $\bardelphitwo$ is
equal to the mean phase noise power $\meanP$. Since $\sigma_{\phi} =
\sqrt{\bardelphitwo}$, we have that $\sigma_{\phi} = \sqrt{\meanP}$,
and so phase jitter is simply an alternative way to represent phase noise 
power. Eq.~(\ref{jitterconv}) is then used to convert phase jitter 
$\sigma_{\phi}$ into timing jitter $\sigma_{t}$. The horizontal axis in
Figure~\ref{fig3} is thus simply an encoding of phase noise power. The largest 
phase noise power value used in the simulations was $\meanP = 0.008$, which 
gives
\begin{displaymath}
\sigma_{t} = \frac{\sqrt{0.008}}{2\pi (10^{9}\,\mathrm{s}^{-1})} 
                  = 14.2\mathrm{ps}.
\end{displaymath}
This corresponds to the right-most data-point in Figure~\ref{fig3}. A similar
conversion of phase noise power was done for the other simulation data-points.
At $\sigma_{t}=5.03\mathrm{ps}$, appropriate for commercially available AWGs, 
Figure~\ref{fig3} indicates that $P_{e}\leq (2.04\pm 
1.80)\times 10^{-5}$. From Table~\ref{table1}, we see that, for ideal control, 
NOC produced a Hadamard gate with $P_{e}\leq 1.04\times 10^{-8}$. As 
anticipated, the NOC performance gains are impacted by phase jitter. 
Figure~\ref{fig3} also shows that if an AWG was available with $\sigma_{t} = 
1.26\mathrm{ps}$, then $P_{e}\leq (9.59\pm 6.94)\times 10^{-7}$, 
which is: (i)~ an order of magnitude reduction in the impact of phase jitter 
compared to $\sigma_{t}=5.03\mathrm{ps}$; and
(ii)~\textit{two orders-of-magnitude\/} less than the target accuracy threshold 
of $10^{-4}$, underscoring the importance of reducing timing jitter in the 
control electronics. We discuss this further below.

In Table~\ref{table5} 
\begin{table*}[!htb]
\centering
\caption{\label{table5}Sensitivity of $TrP$ to timing jitter $\sigma_{t} = 
\sqrt{\meanP}/(2\pi f_{clock})$ for all target gates in the 
universal set $\mathcal{G}_U$. For all gates, the numerical simulations used  
mean noise power $\bar{P}=0.001$, which corresponds to timing jitter 
$\sigma_{t}=5.03 \mathrm{ps}$ for $f_{clock} = 1\mathrm{GHz}$. For each
gate, ten phase noise realizations were generated (see 
Appendix~\ref{appendixNoiseModel}), leading to ten values of the $Tr\, P$
upper bound on the gate error probability $P_{e}\leq Tr\, P$. The 
third column lists, for each gate, the corresponding average $<TrP>$, and uses 
the standard deviation $\sigma (TrP)$ to indicate the spread of $Tr\, P$ about 
the average.\\}
\begin{ruledtabular}
\begin{tabular}{ccc}

  Gate&            Timing-jitter $\sigma_{t}$  & $P_{e}\leq \;\; <TrP> \pm \sigma 
\mathrm{(TrP)}$ with NOC \\
  \hline
Hadamard & $5.03ps$ & $(2.04\pm 1.80)\times 10^{-5}$\\
  NOT& $5.03ps$ & $(2.11\pm 1.64)\times 10^{-5}$\\  
  Modified $\pi/8$& $5.03ps$ & $(2.92\pm 1.96)\times 10^{-5}$\\
  Modified phase& $5.03ps$ & $(3.04\pm 2.16)\times 10^{-5}$\\
  Modified controlled phase& $5.03ps$ & $(5.21\pm 0.00)\times 10^{-5}$\\
\end{tabular}
\end{ruledtabular}
\end{table*}
we display the impact of phase/timing jitter on the NOC performance gains of 
all gates 
in $\calGU$ for timing jitter $\sigma_{t} =5.03\mathrm{ps}$. We see that, even 
with timing jitter at the level found in commercially available AWGs, all gates in 
$\calGU$ have gate error probabilities that are an \textit{order of magnitude 
smaller\/} than the target accuracy threshold value of $10^{-4}$. Notice also
the insensitivity of the two-qubit TRP gate to $5.03\mathrm{ps}$ timing 
jitter. The standard deviation for this gate, $\sigma (TrP) = 5.26\times 
10^{-11}$, is displayed as zero to three significant figures in 
Table ~\ref{table5}.This weak sensitivity to timing jitter is not completely 
surprising given the weak sensitivity of this gate to imprecision in $\lambda$ 
and $\eta_{4}$ that was found in Ref.~\cite{trp6}, and thus to imprecision 
in the twisting of the control field. The critical parameters for this gate are 
$d_{1}$, $d_{4}$, and $c_{4}$ (see Appendix~\ref{appendixD2a}).

In Table~\ref{table6}
 \begin{table*}[!htb]
\centering
 \caption{\label{table6}Sensitivity of $TrP$ to timing jitter $\sigma_{t} = 
\sqrt{\meanP}/(2\pi f_{clock})$ for all target gates in the 
universal set $\mathcal{G}_U$. For all gates, the numerical simulations used  
mean noise power $\bar{P}=6.25\times 10^{-5}$, which corresponds to timing 
jitter $\sigma_{t}=1.26 \mathrm{ps}$ for $f_{clock} = 1\mathrm{GHz}$. For 
each gate, ten phase noise realizations were generated (see 
Appendix~\ref{appendixNoiseModel}), leading to ten values of the $Tr\, P$
upper bound on the gate error probability $P_{e}\leq Tr\, P$. The third column 
lists, for each gate, the corresponding average $<TrP>$, and uses 
the standard deviation $\sigma (TrP)$ to indicate the spread of $Tr\, P$ about 
the average.\\}
 \begin{ruledtabular}
 \begin{tabular}{ccc}
  Gate &   Timing-jitter $\sigma_{t}$ & $P_{e}\leq \;\; <TrP> \pm \sigma 
\mathrm{(TrP)}$ with NOC \\
  \hline
  Hadamard & $1.26ps$ & $(9.59\pm 6.94)\times 10^{-7}$\\
 Modified $\pi/8$& $1.26ps$ & $(1.24\pm 1.04)\times 10^{-6}$\\
  NOT& $1.26ps$ & $(1.82\pm 1.14)\times 10^{-6}$\\
  Modified phase& $1.26ps$ & $(1.92\pm 1.57)\times 10^{-6}$\\
  Modified controlled phase& $1.26ps$ & $(5.21\pm 0.00)\times 10^{-5}$\\
  \end{tabular}
\end{ruledtabular}
\end{table*}
we display the impact of phase/timing jitter on the NOC performance gains
of all gates in $\calGU$ for timing jitter $\sigma_{t}=1.26\mathrm{ps}$. We
see that the gate error probability for the one-qubit gates is reduced by an
order-of-magnitude ($P_{e} \sim 10^{-5}\rightarrow 10^{-6}$) compared to the
error probability at $\sigma_{t} =5.03\mathrm{ps}$. The two-qubit gate 
error probability is unchanged at $P_{e}=5.21\times 10^{-5}$, although its
standard
deviation is now $\sigma (TrP) = 4.24\times 10^{-14}$. Thus reducing timing 
jitter by a factor of $5$ produces one-qubit gates whose error probability is 
\textit{two orders-of-magnitude\/} smaller than the target accuracy threshold
of $10^{-4}$. For a threshold $P_{a}\sim 10^{-3}$ appropriate for surface and
color quantum error correcting codes, \textit{all gates\/} in $\calGU$ operate 
$2$--$3$
\textit{orders-of-magnitude\/} below threshold at $\sigma_{t} = 1.26
\mathrm{ps}$. Thus, for AWGs operating at this reduced level of timing jitter, 
the impact of phase/timing jitter on the NOC performance gains is greatly 
mitigated.

Lastly, note that for starting gates whose good performance is not due to 
quantum interference, phase jitter may have less impact on the NOC performance 
gains than for the TRP gates examined here.

\section{Summary}
\label{sec4}

In this paper we have shown how neighboring optimal control (NOC) theory 
can be used to improve the performance of a good quantum gate. We illustrated 
the NOC theoretical framework by using it to improve the performance
of all gates in a universal set of quantum gates produced using a type of
non-adiabatic rapid passage that has been studied in the literature
\cite{trp1,trp2,trp3,trp4,trp5,trp6,trp7,trp8}. We stress that the NOC
approach introduced here is not limited to this family of starting gates---any 
other good quantum gate, or set of gates, could serve as input for the 
method. For ideal control (see Table~\ref{table1}), the improvements are 
substantial : (i)~for all one-qubit gates in the universal set, the gate error 
probabilities were reduced by \textit{four orders-of-magnitude\/} ($10^{-4}
\rightarrow 10^{-8}$); and (ii)~for the two-qubit gate in the set, by 
\textit{two orders-of-magnitude\/} ($10^{-3}\rightarrow 10^{-5}$). We 
examined the bandwidth required to implement the ideal controls and showed
that for gate times $1\mu\mathrm{s} \leq T \leq 5\mu\mathrm{s}$, the bandwidth 
$\Delta f$ for all gates was in the range $130\mathrm{MHz}\leq \Delta f \leq 820
\mathrm{MHz}$, which is well within the capabilities of commercially available 
arbitrary waveform generators. We examined the robustness of these performance 
improvements to two important sources of non-ideal control: (i)~control 
parameters with finite precision; and (ii)~timing/phase jitter resulting for 
clock errors in the control electronics. We showed (see Section~\ref{sec3c1}
and Appendix~\ref{appendixD2a}) that the NOC performance 
gains require arbitrary waveform generators with $14$-bit ($17$-bit) 
vertical resolution for the one-qubit (two-qubit) gates. We also showed 
(see Section~\ref{sec3c2} and Appendix~\ref{appendixD2b}) that 
timing/phase jitter can significantly impact the NOC performance gains. We 
showed that for $5\mathrm{ps}$ timing jitter (comparable to that in commerically
available AWGs), the gate error probability satisfies $P_{e}\sim 10^{-5}$ for 
all the gates in the universal set, an order-of-magnitude lower than the 
accuracy threshold target value of $10^{-4}$. Finally, we showed 
(Section~\ref{sec3c2}) that if timing jitter can be reduced to $\sigma_{t} =
1.26\mathrm{ps}$, the error probability for all one-qubit gates in $\calGU$  
drops to $P_{e}\sim 10^{-6}$, while the two-qubit gate error probability
remains unchanged at $5.21\times 10^{-5}$. All gates thus operate with an
error probability $1$--$2$ \textit{orders-of-magnitude\/} below the target 
threshold of $10^{-4}$. Although we have focused on a target accuracy threshold 
$P_{a}= 10^{-4}$ throughout this paper, we note that for surface and color 
quantum error correcting codes, the accuracy threshold satisfies $P_{a}\sim 
10^{-3}$ \cite{homcodes1,homcodes2,homcodes3,homcodes4,homcodes5}. 
For these codes, the NOC improved gates \textit{all\/} operate $2$--$3$ 
\textit{orders-of-magnitude\/} below threshold, even for non-ideal control. The
availability of a universal set of quantum gates operating so far below 
threshold would have a significant impact on efforts to realize fault-tolerant 
quantum computing as it would greatly reduce the resources 
needed to implement such a computation. It is hoped that the NOC gate 
performance improvements found in this paper might encourage an attempt 
to produce these high-fidelity gates experimentally.

We close by noting that we have assumed throughout this paper that the qubit
longitudinal ($T_{1}$) and transverse ($T_{2})$ relaxation times are long 
compared to the gate operation time $T_{gate}$. This assumption is essential 
for any discussion of fault-tolerant quantum computing and error correction 
as it insures that the qubit state does not decohere away before the 
error-syndrome extraction circuit can be applied, and likely errors 
identified. When $T_{1},T_{2}\gg T_{gate}$, control imperfections may be 
anticipated to be the primary source of errors during a gate operation, and the 
qubit environment a secondary source. On the other hand, when 
$T_{1},T_{2} \alt T_{gate}$, the qubits are of sufficiently poor quality that 
errors from the qubit environment can be expected to be (at least) as bad as 
the types of errors we have examined in this paper. Our NOC strategy 
for improving a good quantum gate does not remove the need for high-quality 
qubits as the object of these gate operations.

\begin{acknowledgments}
We thank:~C. Tahan for providing the computing resources used to obtain the
simulation results reported in this paper; and F. Wellstood, V. Yakovenko, and 
R. Li for valuable discussions. F.~G.\ thanks T. Howell~III for continued 
support.\\
\end{acknowledgments}

\appendix

\section{Twisted rapid passage - a few more necessary results}
\label{appendixTRP}

We illustrated the general theory developed in Section~\ref{sec2} by using it
in Section~\ref{sec3} and Appendix~\ref{appendixRemainingResults} to improve
the performance of a universal set of quantum gates implemented using 
a form of non-adiabatic rapid passage known as twisted rapid passage (TRP)
\cite{trp1}-\cite{trp8}. In Section~\ref{sec3a} we provided a brief introduction
to TRP. In this Appendix we complete our review of TRP. 
Appendix~\ref{appendixA1} presents the derivation of the dimensionless 
one- and two-qubit Hamiltonians used to drive the quantum gates produced 
using TRP. Appendix~\ref{appendixA2} derives an expression for the gate error
probability, as well as a convenient upper bound for it. Finally, for the 
reader's convenience, Appendix~\ref{appendixA3} collects previous 
results for the TRP sweep parameters, gate error probabilities and fidelities 
for the TRP-generated universal set of quantum gate studied in 
Refs.~\cite{trp6} and \cite{trp8}. It also provides the TRP approximate
gates $U_{0}(t=T)$ for each gate in $\calGU$. These gates serve 
as the good starting gates that are improved using neighboring optimal 
control. We stress that this approach to improving a good quantum gate (or 
set of gates) is not limited to this TRP-generated family of gates. Any good 
gate could provide the starting point for the method. 

\subsection{One- and two-qubit Hamiltonians}
\label{appendixA1}

\textit{(a)}~For the one-qubit gates studied in this paper, the qubit is assumed
to couple to an external control field $\bff (t)$ through the 
Zeeman-interaction,
\begin{equation}
H_{0}^{1}(t) = -\bldsig\cdot\bff (t) ,
\label{appendixZeeman}
\end{equation}
where $\bff (t)$ has the TRP profile,
\begin{equation}
\bff (t) = at\bfhatz + b\cos\phi_{4} (t)\bfhatx + b\sin\phi_{4} (t)\bfhaty ,
\label{appendixTRPprof}
\end{equation}
and for quartic twist, $\phi_{4}(t) = (1/2)Bt^{4}$ with $-T/2\leq t \leq T/2$. 
The Schrodinger equation for the propagator $U(t,-T/2)$ is
\begin{equation}
i\hbar\frac{dU(t)}{dt} = \left[ -at\sigma_{z} - b\cos\phi_{4}(t)\sigma_{x}
                                      -b\sin\phi_{4}(t)\sigma_{y}\right] U(t),
\label{onequbitScheq}
\end{equation}
where we have suppressed the $-T/2$ dependence in $U(t,-T/2)$. It proves useful
to express Eq.~(\ref{onequbitScheq}) in dimensionless form. To that end we
define: (i)~the dimensionless time $\tau = (a/b)t$; (ii)~the dimensionless 
inversion rate $\lambda = \hbar a/b^{2}$; and (iii)~the dimensionless twist 
strength $\eta_{4} = \hbar Bb^{2}/a^{3}$. In terms of these parameters, 
Eq.~(\ref{onequbitScheq}) becomes
\begin{equation}
i\frac{dU(\tau )}{d\tau} = H_{0}^{1}(\tau ) U(\tau),
\label{dim1qbtScheq}
\end{equation}
where the dimensionless one-qubit Hamiltonian is
\begin{equation}
H_{0}^{1}(\tau ) = \frac{1}{\lambda}\left[ -\tau\sigma_{z} -\cos\phi_{4}(\tau )
                            \sigma_{x} -\sin\phi_{4}(\tau )\sigma_{y}\right],
\label{dim1qbtHam}
\end{equation}
and $\phi_{4}(\tau ) = (\eta_{4}/2\lambda )\tau^{4}$. This is the nominal
one-qubit Hamiltonian discussed in Section~\ref{sec3a3} that drives the 
numerical simulation of all one-qubit gates considered in this paper. 

\textit{(b)}~Next we derive the dimensionless nominal two-qubit Hamiltonian 
$H_{0}^{2}(\tau )$ discussed in Section~\ref{sec3a3} and which drives the 
numerical simulations of the two-qubit modified controlled phase gate. Although 
a more general discussion is possible, it proves convenient to adopt the 
language of NMR which was the original experimental setting for TRP 
\cite{trp2,zwan}.

The two-qubit Hamiltonian contains terms that Zeeman-couple each qubit to an
external control field $\bff (t)$, and an Ising interaction term that couples the 
two qubits. Note that alternative two-qubit interactions can easily be considered 
by straightforward modification of the following arguments. Our starting point 
is thus the Hamiltonian
\begin{equation}
\frac{\overline{H}_{\: 0}^{ \: 2}(t)}{\hbar} = -\frac{1}{2}\sum_{i=1}^{2}\,
                    \gamma_{i}\,\bldsig^{i}\cdot\bff (t) - \frac{\pi}{2}\, J\, 
                          \sigma_{z}^{1}\sigma_{z}^{2} ,
\label{Ham1}
\end{equation} 
where $\gamma_{i}$ is the gyromagnetic ratio for qubit~$i$, and $J$ is the
Ising interaction coupling constant. In the lab frame, $\bff (t)$ has a
static component $B_{0}\,\bfhatz$ and a time-varying component $2B_{rf}\cos
\phi_{rf}(t)\,\bfhatx$. In the rotating wave approximation $\bff (t)$ reduces
to
\begin{equation}
\bff (t) = B_{0}\,\bfhatz + B_{rf}\cos\phi_{rf}(t)\,\bfhatx -B_{rf}\sin
            \phi_{rf}(t)\,\bfhaty .
\label{Brwa}
\end{equation}
Introducing $\omega_{i}=\gamma_{i}B_{0}$ and $\omega_{i}^{rf}=\gamma_{i}
B_{rf}$ ($i=1,2$), and inserting Eq.~(\ref{Brwa}) into Eq.~(\ref{Ham1}) gives
\begin{eqnarray}
\frac{\overline{H}_{\: 0}^{\: 2}(t)}{\hbar} & = & \sum_{i=1}^{2}\,\left[
         -\frac{\omega_{i}}{2}\sigma_{z}^{i} -  \frac{\omega_{i}^{rf}}{2}
          \left\{\,\cos\phi_{rf}\sigma_{x}^{i}-\sin\phi_{rf}\sigma_{y}^{i}
               \right\} \right] \nonumber\\
 & & \hspace{1in} - \frac{\pi}{2}\, J\, \sigma_{z}^{1} \sigma_{z}^{2} .
\end{eqnarray}
Transformation to the detector frame is done via the unitary operator
\begin{displaymath}
U_{det}(t)=\exp\left[\, (i\phi_{det}(t)/2)\left(\sigma_{z}^{1}+
\sigma_{z}^{2}\right)\,\right] .
\end{displaymath} 
The Hamiltonian in the detector frame is then \cite{abra}
\begin{eqnarray}
\frac{\mbox{\~{H}}_{0}^{2}(t)}{\hbar}  & = &  U^{\dagger}_{det}
   \left( \frac{\overline{H}_{\: 0}^{\: 2}(t)}{\hbar}\right) U_{det}
                          -iU^{\dagger}_{det}\frac{dU_{det}}{dt}  \nonumber\\
 & = & \sum_{i=1}^{2}\left[\left( -\frac{\omega_{i}}{2}+\dot{\phi}_{det}
          \right)\sigma_{z}^{i} \right. \nonumber\\
 & & \hspace{0.1in}  -\frac{\omega_{i}^{rf}}{2}\left\{\cos\left(
           \phi_{det}-\phi_{rf}\right)\sigma_{x}^{i}\right. \nonumber \\
  & & \hspace{0.7in} \left.\left. +\sin\left(\phi_{det}-
            \phi_{rf}\right)\sigma_{y}^{i}\right\}\right]\nonumber\\ 
 & & \hspace{1in} -\frac{\pi}{2}\, J\, \sigma_{z}^{1}\sigma_{z}^{2} .
\label{Ham2}
\end{eqnarray}
As explained in Refs.~\cite{trp2,zwan}, to produce a TRP sweep in the detector
frame it is necessary to sweep $\dot{\phi}_{det}$ and $\dot{\phi}_{rf}$ 
through a Larmor resonance frequency. We choose (somewhat arbitrarily) to 
sweep through the Larmor frequency $\omega_{2}$: 
\begin{eqnarray}
\dot{\phi}_{det} & = & \omega_{2} +\frac{2at}{\hbar} + \Delta \nonumber\\
\dot{\phi}_{rf} & = & \dot{\phi}_{det} - \dot{\phi}_{4} .
\label{twoqbtdefs}
\end{eqnarray}
Here $\phi_{4}(t) = (1/2)Bt^{4}$ is the twist profile for quartic TRP, and we 
have introduced a frequency shift parameter $\Delta$ whose value is determined 
by the sweep parameter optimization procedure described in Ref.~\cite{trp6}.
Inserting Eqs.~(\ref{twoqbtdefs}) into Eq.~(\ref{Ham2}),
and introducing $\delta\omega = \omega_{1}-\omega_{2}$ and $b_{i}=\hbar
\omega_{i}^{rf}/2$ ($i=1,2$), we find
\begin{eqnarray}
\frac{\mbox{\~{H}}_{0}^{2}(t)}{\hbar}  & = & 
 \left[ -
         \frac{(\delta\omega +\Delta )}{2} + \frac{at}{\hbar}\right]
                  \sigma_{z}^{1}
         +\left[ - \frac{\Delta}{2}+\frac{at}{\hbar}\right]
         \sigma_{z}^{2}  \nonumber \\
  & & \hspace{0.3in}   -\frac{b_{1}}{\hbar}\left[\,\cos\phi_{4}\,\sigma_{x}^{1}
            +\sin\phi_{4}\, \sigma_{y}^{1}\,\right] \nonumber \\
  & & \hspace{0.6in}  -  \frac{b_{2}}{\hbar}\left[\,
               \cos\phi_{4}\,\sigma_{x}^{2}\sin\phi_{4}\,\sigma_{y}^{2}\, \right] 
                      \nonumber \\
  & & \hspace{1in} -\frac{\pi}{2}\, J\,\sigma_{z}^{1}\sigma_{z}^{2} .
\label{Ham3}
\end{eqnarray}
We see that both qubits are acted on  by a quartic TRP sweep in the detector
frame. In keeping with our earlier choice of sweeping through the 
Larmor resonance of the second qubit, we use $b_{2}$ in the definitions of the 
dimensionless time $\tau$, inversion rate $\lambda$, and twist strength 
$\eta_{\scriptscriptstyle 4}$:
\begin{eqnarray}
\tau & = & \left(\frac{a}{b_{2}}\right)\, t \label{twoqbttdef}\\
\lambda & = & \frac{\hbar a}{\left( b_{2}\right)^{2}} \\
\eta_{\scriptscriptstyle 4} & = & \left(\frac{\hbar B}{a^{3}}\right)\,
        \left( b_{2}\right)^{2} .
\label{twoqbte4def}
\end{eqnarray}
Since $\mbox{\~{H}}_{0}^{2}(t)/\hbar$ has units of inverse-time, and $b_{2}/a$ 
has units of time (Eq.~(\ref{twoqbttdef})), multiplying Eq.~(\ref{Ham3}) by
$b_{2}/a$ and using Eqs.~(\ref{twoqbttdef})--(\ref{twoqbte4def}) gives the 
dimensionless two-qubit Hamiltonian $\mbox{\~{H}}_{0}^{2}(\tau )$:
\begin{eqnarray}
\mbox{\~{H}}_{2}(\tau ) & = & \left[ -\frac{(d_{1}+d_{2})}{2}+
                               \frac{\tau}{\lambda}\right]\,\sigma_{z}^{1}
+\left[ -\frac{d_{2}}{2} + \frac{\tau}{\lambda}\right]\,
        \sigma_{z}^{2}  \nonumber\\
 & & \hspace{0.25in}  -\frac{d_{3}}{\lambda}\left[\,\cos\phi_{4}\,
              \sigma_{x}^{1} +\sin\phi_{4}\,\sigma_{y}^{1}\,\right] \nonumber\\
  & & \hspace{0.5in} 
        -\frac{1}{\lambda}\left[\,\cos\phi_{4}\,\sigma_{x}^{2}+\sin\phi_{4}\,
          \sigma_{y}^{2}\,\right] \nonumber\\
  & & \hspace{1in} -\frac{\pi}{2}\, d_{4}\,\sigma_{z}^{1}\sigma_{z}^{2} ,
\end{eqnarray}
where
\begin{eqnarray}
d_{1} & = & \left(\frac{\delta\omega}{a}\right)\, b_{2} \nonumber\\
d_{2} & = & \left(\frac{\Delta}{a}\right)\, b_{2} \nonumber\\
d_{3} & = & \frac{b_{1}}{b_{2}} \nonumber \\
d_{4} & = & \left(\frac{J}{a}\right)\, b_{2} .
\label{ddefs}
\end{eqnarray}
As noted in Section~\ref{sec3a3}, $\mbox{\~{H}}_{2}(\tau )$ has a degeneracy
in the resonance frequency of the energy level pairs ($E_{1}\leftrightarrow
E_{2}$) and ($E_{3}\leftrightarrow E_{4}$). To break this degeneracy we add 
the term
\begin{equation}
\Delta H = c_{4}\, |E_{4}(\tau )\rangle\langle E_{4}(\tau )|
\end{equation} 
to $\mbox{\~{H}}_{0}^{2}(\tau )$, where $|E_{4}(\tau )\rangle$ is the 
instantaneous energy eigenstate of $\mbox{\~{H}}_{0}^{2}(\tau )$ with 
eigenvalue $E_{4}(\tau )$. Our final Hamiltonian is then
\begin{equation}
H_{0}^{2}(\tau ) = \mbox{\~{H}}_{0}^{2}(\tau ) + \Delta H
\end{equation}
which is the Hamiltonian given in Eq.~(\ref{twoqbtHam}). We see that 
$H_{0}^{2}(\tau )$ depends on the TRP sweep parameters ($\lambda$, 
$\eta_{\scriptscriptstyle 4}$), as well as on the parameters ($d_{1},\ldots ,
d_{4}$) and $c_{4}$. From Eq.~(\ref{ddefs}) we see that $d_{1}$, $d_{2}$,
$d_{3}$, and $d_{4}$ are the dimensionless versions of, respectively, the
Larmor frequency difference $\delta\omega = \omega_{1}-\omega_{2}$, the
frequency shift parameter $\Delta$, the ratio $b_{1}/b_{2}=\gamma_{1}/
\gamma_{2}$, and the Ising coupling constant $J$.

For a derivation of the one-qubit TRP Hamiltonian (Eq.~(\ref{appendixZeeman})) 
based on an NMR experimental implementation, see the Appendix of 
Ref.~\cite{trp1}.

\subsection{Gate error probability}
\label{appendixA2}

The following argument is for an $N$-dimensional Hilbert space. As in 
Section~\ref{sec3a3}, let $U_{a}$ denote the actual unitary operation 
produced by a given set of TRP sweep parameters and $U_{tgt}$ a target unitary 
operation we would like TRP to approximate as closely as possible. Introducing 
the operators $D=U_{a}-U_{tgt}$ and $P=D^{\dagger}D$, and the normalized state 
$|\psi\rangle$, we define $|\psi_{a}\rangle = U_{a}|\psi\rangle$ and 
$|\psi_{tgt}\rangle =U_{tgt}|\psi\rangle$. Now choose an orthonormal basis 
$|i\rangle$ ($i=1,\ldots ,N$) such that $|1\rangle\equiv |\psi_{tgt}\rangle$ 
and define the state $|\xi_{\psi}\rangle$ via
\begin{eqnarray}
|\psi_{a}\rangle & = & |\psi_{tgt}\rangle + |\xi_{\psi}\rangle
   \label{eq231} \\
 & = & |1\rangle + |\xi_{\psi}\rangle \hspace{0.1in} . \label{eq232}
\end{eqnarray}
Inserting $|\xi_{\psi}\rangle = \sum_{i=1}^{N}\, e_{i}|i\rangle$ into 
eq.~(\ref{eq232}) gives
\begin{equation}
|\psi_{a}\rangle = \left( 1+e_{1}\right)|1\rangle + \sum_{i\neq 1}\, e_{i}
 |i\rangle  \hspace{0.1in} .
\label{eq233}
\end{equation}
Since $|\psi_{tgt}\rangle = |1\rangle$ is the target state, it is clear from
Eq.~(\ref{eq233}) that the error probability $P_{e}(\psi )$ for $U_{a}$ 
(i.~e.~TRP) is
\begin{eqnarray}
P_{e}(\psi ) & = & \sum_{i\neq 1}\, |e_{i}|^{2} \hspace{0.1in} .
 \label{eq235}
\end{eqnarray}
We define the error probability $P_{e}$ for the TRP gate to be
\begin{equation}
P_{e} \equiv \max_{\scriptstyle |\psi\rangle}\, P_{e}(\psi ) 
\hspace{0.1in} .  \label{eq236}
\end{equation}
From Eq.~(\ref{eq231}),
\begin{displaymath}
|\xi_{\psi}\rangle = D|\psi\rangle
\end{displaymath}
and
\begin{eqnarray}
\langle\xi_{\psi}|\xi_{\psi}\rangle & = & \langle\psi |D^{\dagger}D|
     \psi\rangle \nonumber \\
 & = & Tr\rho_{\psi}P \hspace{0.1in} , \label{eq23650}
\end{eqnarray}
where $\rho_{\psi} = |\psi\rangle\langle\psi |$. On the other hand,
\begin{eqnarray}
\langle\xi_{\psi}|\xi_{\psi}\rangle & = & \sum_{i=1}^{N}\, |e_{i}|^{2}
 \nonumber \\
 & = & |e_{1}|^{2} + P_{e}(\psi ) \hspace{0.1in} . \label{eq23675}
\end{eqnarray}
Combining Eqs.~(\ref{eq23650}) and (\ref{eq23675}) gives
\begin{eqnarray*}
P_{e}(\psi ) & = & \langle\xi_{\psi}|\xi_{\psi}\rangle - |e_{1}|^{2} \\
 & \leq & \langle\xi_{\psi}|\xi_{\psi}\rangle = Tr\rho_{\psi} P 
 \hspace{0.1in} .
\end{eqnarray*}
Since $P=D^{\dagger}D$ is
Hermitian it can be diagonalized: $P=O^{\dagger}d\, O$ and $d=diag(d_{1},
\ldots , d_{N})$. Thus
\begin{displaymath}
P_{e}(\psi ) \leq Tr\,\overline{\rho}_{\psi}d \hspace{0.1in} ,
\end{displaymath}
where $\overline{\rho}_{\psi} = O\rho_{\psi}O^{\dagger}$. Let $d_{\ast} =
\max (d_{1}, \ldots ,d_{N})$, then direct evaluation of the trace gives
\begin{eqnarray*}
Tr\,\overline{\rho}_{\psi}d & = & \sum_{i=1}^{N} d_{i}
\left(\overline{\rho}_{\psi}\right)_{ii} \\
 & \leq & \sum_{i=1}^{N} d_{\ast}\left(\overline{\rho}_{\psi}\right)_{ii}
  = d_{\ast}\, Tr\,\overline{\rho}_{\psi} = d_{\ast} \hspace{0.1in} ,
\end{eqnarray*}
where we have used that $Tr\,\overline{\rho}_{\psi}=1$. Thus $P_{e}(\psi )
\leq d_{\ast}$ for \textit{all\/} states $|\psi\rangle$. 
From Eq.~(\ref{eq236}), it follows that
\begin{equation}
P_{e}\leq d_{\ast} \hspace{0.1in} ,
\label{eq237}
\end{equation}
so that the largest eigenvalue $d_{\ast}$ of $P$ is an upper bound for the 
gate error probability $P_{e}$. Finally, notice that $P=D^{\dagger}D$ is a 
positive operator so that $d_{i}\geq 0$ for $i=1,\ldots , N$. Thus $d_{\ast}
\leq Tr\, P$ and so
\begin{equation}
P_{e}\leq d_{\ast}\leq Tr\, P \hspace{0.1in} .
\label{eq238}
\end{equation}
Although $Tr\, P$ need not be as tight an upper bound on $P_{e}$ as $d_{\ast}$,
it is much easier to calculate and so is more convenient than $d_{\ast}$ for 
use in the numerical simulations carried out in this paper. 

\subsection{Nominal gates}
\label{appendixA3}

 The nominal quantum gates whose performance is to be improved through 
neighboring optimal control are the set of one-qubit gates examined in 
Ref.~\cite{trp8}, and the two-qubit modified controlled phase gate 
studied in Ref.~\cite{trp6}. As these papers showed, these gates provide
a good approximation to the universal quantum gate set $\calGU$ introduced 
in Section~\ref{sec3a2}. For the reader's convenience we reproduce in this 
subsection the main results of these papers  which, for each gate, include: 
(i)~the control parameters used to produce the approximate gate; (ii)~the
$Tr\, P$ upper bound on its gate error probability $P_{e}$; and (iii)~its gate 
fidelity $\calF$. These results are collected in Tables~\ref{tableapp1} and 
\ref{tableapp2} below. We also include the TRP-generated unitary gate 
$U_{0}(\tau = \tau_{0}/2)$ for each quantum gate in $\calGU$.

\textbf{One-qubit gates:} As was shown in Section~\ref{sec3a3} and 
Appendix~\ref{appendixA1}, the parameters $\lambda$, $\eta_{4}$, and 
$\tau_{0} = aT/b$ fix the TRP control field $\bff_{0}(\tau )$ that implements a 
particular nominal one-qubit gate. In all our one-qubit simulations $\tau_{0}=
160$ \cite{foot5}. Table~\ref{tableapp1}
\begin{table*}[!htb]
\centering
 \caption{\label{tableapp1} The nominal one-qubit gates used in this paper are
those studied in Ref.~\cite{trp8}. For the reader's convenience, for each gate, 
we tabulate the control parameter values and gate performance reported in that 
work. The TRP sweep parameter values listed for $\lambda$ and $\eta_{4}$ were 
found using the downhill simplex optimization algorithm; the $TrP$ upper bound 
on the gate error probability (see Eq.~(\ref{eq238})) was found using numerical 
simulation of the one-qubit Schrodinger dynamics; and the gate fidelity 
$\mathcal{F}$ follows from $Tr\, P$ (see Sec.~\ref{sec3a3}). The dimensionless 
inversion time $\tau_0 = 160$.\\}
 \begin{ruledtabular}
 \begin{tabular}{c|cccc}
   Gate & $\lambda$ & $\eta_4$ &$TrP$ & $\mathcal{F}$\\ \hline
   NOT              & $6.965$ & $2.189\times 10^{-4}$ & $6.27\times 10^{-5}$ 
                & $0.99998$ \\
   Hadamard         & $7.820$ & $1.792\times 10^{-4}$ & $1.12\times 10^{-4}$ 
               & $0.99997$ \\
   Modified $\pi/8$ & $8.465$ & $1.675\times 10^{-4}$ & $2.13\times 10^{-4}$ 
              & $0.99995$ \\
   Modified phase   & $8.073$ & $1.666\times 10^{-4}$ & $4.62\times 10^{-4}$ 
             & $0.99988$ \\
 \end{tabular}
 \end{ruledtabular}
\end{table*}
lists the one-qubit target gates, and for each gate, the TRP control parameters 
that produce a good approximation $U_{a}$ to it. Column~$3$ gives the upper 
bound $Tr\, P$ on the gate error probability $P_{e}$, and column~$4$ gives the
gate fidelity $\calF$ (see Section~\ref{sec3a3}). Ref.~\cite{trp8} describes the
optimization procedure used to determine the control parameter values appearing
in the Table.

Finally, we include the unitary gates produced by the TRP sweep parameters 
listed in Table~\ref{tableapp1}.\\

\noindent\textit{(1)} For the NOT gate, the TRP-generated unitary is:
\begin{displaymath}
U_{NOT} = \left(  \begin{array}{cc}
                                   -0.0014 + 0.0000\, i &  1.0000 + 0.0054\, i\\
                                   1.0000 - 0.0054\, i & 0.0014 + 0.0000\, i
                             \end{array}  \right) .
\end{displaymath}
With $U_{0}(\tau =\tau_{0}/2) = U_{NOT}$ and $U_{tgt} = \sigma_{x}$, we find 
that
\begin{displaymath}
Tr\,\left[ U^{\dagger}_{0}(\tau_{0}/2)U_{tgt}\right] = 2 + 3.2000\times 10^{-5} .
\end{displaymath}
Recall that $\delta\beta = i\left[ U^{\dagger}_{0}(\tau_{0}/2)U_{tgt} - I
\right]$.
Using the max-norm $\| U\| = \max_{i,j} |U_{ij}|$, we can show that $\|\delta
\beta\| = 0.0054$. This sets the scale for small quantities introduced in 
Section~\ref{sec2}: $\Delta = \|\delta\beta\|$. Thus $\Delta^{2} = 2.92\times 
10^{-5}$, and so we see that
\begin{displaymath}
Tr\,\left[ U^{\dagger}_{0}(\tau_{0}/2)U_{tgt}\right] = 2 + \calO (\Delta^{2}) .
\end{displaymath}

\noindent\textit{(2)} For the Hadamard gate, the TRP-generated unitary is:
\begin{displaymath}
U_{H} = \left(  \begin{array}{cc}
                                   0.7112 + 0.0000\, i &  0.7030 - 0.0016\, i\\
                                   0.7030 + 0.0016\, i & -0.7112 + 0.0000\, i
                             \end{array}  \right) .
\end{displaymath}
With $U_{0}(\tau =\tau_{0}/2) = U_{H}$ and $U_{tgt} = (1/\sqrt{2})\left(
\sigma_{x}+\sigma_{z}\right)$, we find that
\begin{displaymath}
Tr\,\left[ U^{\dagger}_{0}(\tau_{0}/2)U_{tgt}\right] = 2 + 6.7615\times 
10^{-5} .
\end{displaymath}
Here $\|\delta\beta\| = 0.0081$ and so $\Delta^{2} = 6.561\times 10^{-5}$. 
Thus we see that
\begin{displaymath}
Tr\,\left[ U^{\dagger}_{0}(\tau_{0}/2)U_{tgt}\right] = 2 + \calO (\Delta^{2}) .
\end{displaymath}

\noindent\textit{(3)} For the modified $\pi /8$ gate, the TRP-generated unitary 
is:
\begin{displaymath}
V_{\pi /8} = \left(  \begin{array}{cc}
                                   -0.0061 + 0.0000\, i &  0.9204 + 0.3910\, i\\
                                   0.9204 - 0.3910\, i & 0.0061 + 0.0000\, i
                             \end{array}  \right) .
\end{displaymath}
With $U_{0}(\tau =\tau_{0}/2) = V_{\pi /8}$ and $U_{tgt} = \cos (\pi /8)
\sigma_{x}-\sin (\pi /8)\sigma_{y}$, we find that
\begin{displaymath}
Tr\,\left[ U^{\dagger}_{0}(\tau_{0}/2)U_{tgt}\right] = 2 + 1.2034\times 
10^{-4} .
\end{displaymath}
Here $\|\delta\beta\| = 0.0091$ and so $\Delta^{2} = 8.2810\times 10^{-5}$. 
Thus we see that
\begin{displaymath}
Tr\,\left[ U^{\dagger}_{0}(\tau_{0}/2)U_{tgt}\right] = 2 + \calO (\Delta^{2}) .
\end{displaymath}

\noindent\textit{(4)} For the modified phase gate, the TRP-generated unitary 
is:
\begin{displaymath}
V_{p} = \left(  \begin{array}{cc}
                                   0.0051 + 0.0000\, i &  0.7171 + 0.6969\, i\\
                                   0.7171 - 0.6969\, i & -0.0051 + 0.0000\, i
                             \end{array}  \right) .
\end{displaymath}
With $U_{0}(\tau =\tau_{0}/2) = V_{p}$ and $U_{tgt} = (1/\sqrt{2})\left(
\sigma_{x}-\sigma_{y}\right)$, we find that
\begin{displaymath}
Tr\,\left[ U^{\dagger}_{0}(\tau_{0}/2)U_{tgt}\right] = 2 + 2.3131\times 
10^{-4} .
\end{displaymath}
Here $\|\delta\beta\| = 0.0143$ and so $\Delta^{2} = 2.0449\times 10^{-4}$. 
Thus we see that
\begin{displaymath}
Tr\,\left[ U^{\dagger}_{0}(\tau_{0}/2)U_{tgt}\right] = 2 + \calO (\Delta^{2}) .
\end{displaymath}

\textbf{Two-qubit gate:} As seen in Appendix~\ref{appendixA1}, the two-qubit 
nominal Hamiltonian $H_{0}^{2}(\tau )$ used to produce a good approximation 
to the two-qubit modified controlled phase gate $V_{cp}$ is specified by the 
TRP sweep parameters $\lambda$, $\eta_{4}$, and $\tau_{0}$, as well as the 
parameters $d_{1},\ldots , d_{4}$ and $c_{4}$. All two-qubit simulations used 
$\tau_{0} = 120$. Table~\ref{tableapp2}
\begin{table*}[!htb]
\centering
 \caption{\label{tableapp2} The nominal two-qubit gate used in this paper is
the modified controlled phase gate $V_{cp}$ studied in Ref.~\cite{trp6}. For 
the reader's convenience, we tabulate the control parameter values and gate 
performance reported in that work. The control parameter values listed for 
$\lambda$, $\eta_{4}$, $d_{1}, \ldots , d_{4}$, and $c_{4}$ were found using 
simulated annealing; the $TrP$ upper bound on the gate error probability (see 
Eq.~(\ref{eq238})) was found using numerical simulation of the one-qubit 
Schrodinger dynamics; and the gate fidelity $\mathcal{F}$ follows from 
$Tr\, P$ (see Sec.~\ref{sec3a3}). The dimensionless inversion time $\tau_0 
= 120$.\\}
 \begin{ruledtabular}
 \begin{tabular}{ccccccccc}
  $\lambda$            &$\eta_{4}$ & $d_1$ & $d_2$& $d_3$ & $d_4$ & $c_4$ 
&$TrP$ &$\mathcal{F}$\\
  \hline
  $5.1$ & $2.4\times 10^{-4}$  & 11.702 & -2.6 & -0.41 & 6.6650 & 5.0003 
          & $1.27\times 10^{-3}$ & $0.99984$\\
  \end{tabular}
\end{ruledtabular}
\end{table*}
lists the values for the remaining control parameters; the $Tr\, P$ 
upper bound on the gate error probability $P_{e}$; and the gate fidelity
$\calF$. Ref.~\cite{trp6} describes the optimization procedure used to
determine the control parameter values appearing in the Table.

For the modified controlled-phase gate, the TRP-generated unitary is:
\begin{displaymath}
Re(V_{cp}) = \left(   
               \begin{array}{cccc}
                  0.9998 & 0.0155 & 0.0041 & 0.0028 \\
                 -0.0154 & 0.9997 & -0.0003 & 0.0021 \\
                  0.0042 & -0.0002 & -0.9999 & -0.0038 \\
                -0.0026 & -0.0021 & -0.0037 & 0.9999\\
              \end{array} \right) ;
\end{displaymath}
\begin{displaymath}
Im(V_{cp}) = \left(   
               \begin{array}{cccc}
                  0.0052 & -0.0108 & -0.0031 & -0.0017 \\
                 -0.0109 & 0.0064 & -0.0084 & 0.0068 \\
                  0.0030 & 0.0084 & 0.0060 & -0.0079 \\
                -0.0018 & 0.0068 & 0.0079 & 0.0026\\
              \end{array} \right) .
\end{displaymath}

Finally, it is worth noting that Ref.~\cite{trp7} improved the performance of 
the modified controlled phase gate presented in Ref.~\cite{trp6} by interleaving
a dynamical decoupling pulse sequence with the TRP control field. Although this 
complicates the time-dependence of the control field, it leads to an order of
magnitude reduction in $Tr\, P$ ($Tr\, P = 1.27\times 10^{-3} \rightarrow
8.87\times 10^{-5}$), and only requires control parameters with $14$-bit
precision, compared to the $17$-bit precision required in Ref.~\cite{trp6}.
The reader is referred to Ref.~\cite{trp7} for further details. Although 
this new procedure produces a more robust high fidelity gate, the 
price paid is a control field that is much more difficult to implement 
experimentally. For this reason, in this paper, we have used the modified 
controlled phase gate studied in Ref.~\cite{trp6} as our nominal two-qubit 
gate.

\section{Derivation of Eq.~(\ref{bigsurprise})}
\label{appendixB}

In this Appendix we derive Eq.~(\ref{bigsurprise}) which we re-write here for
convenience:
\begin{eqnarray}
\calI & = & \sum_{j=1}^{3}\barG_{j}\left(G^{\dagger}\bldw \right)_{j}
                     \nonumber\\
  & = & \left( \begin{array}{cc}
                         w_{1} - w_{4} & 2w_{3}\\
                         2w_{2} & w_{1} -w_{4}
                     \end{array} \right) ,
\label{bigeqn1}
\end{eqnarray}
where $\bldw$ is a constant vector introduced in Section~\ref{sec2d}. To 
avoid cluttering equations, we suppress the time dependence of all vectors and
matrices throughout this Appendix. We begin in Appendix~\ref{appendixB1} by
introducing a number of definitions aimed at making the flow of the calculation 
of $\calI$ clearer, and then move on to the calculation of $\calI$ in 
Appendix~\ref{appendixB2}.\\

\subsection{Preliminary definitions}
\label{appendixB1}

Our derivation assumes the quantum system of interest is a single qubit whose 
dynamics is driven by the Zeeman Hamiltonian $H = -\bldsig\cdot\bff $. 
Following the development in Section~\ref{sec2a}, for this Hamiltonian, 
$\calG_{j} = -\sigma_{j}$, where the $1,2,3$ components of $\bldsig$ are the 
$x$, $y$, $z$ Pauli matrices, respectively, and
\begin{displaymath}
\barG_{j} = U^{\dagger}_{0}\calG_{j}U_{0} = - U^{\dagger}_{0}\sigma_{j}U_{0},
\end{displaymath}
with
\begin{displaymath}
U_{0} = \left( \begin{array}{cc}
                           U_{11} & U_{12}\\
                          U_{21} & U_{22}
                       \end{array} \right)  = 
 \left(  \begin{array}{cc}
               \vline & \vline \\
               \bldc_{1} & \bldc_{2} \\
              \vline & \vline 
          \end{array} \right).
\end{displaymath}
It follows from the unitarity of $U_{0}$ that $\bldc_{1}$ and $\bldc_{2}$ form 
an orthonormal set: $\bldc_{i}^{\dagger}\bldc_{j} = \delta_{ij}$.

It proves useful to define the vector pairs $(\blde_{1},\blde_{2})$,
$(\bldf_{1},\bldf_{2})$, and $(\bldg_{1},\bldg_{2})$ as follows:
\begin{eqnarray*}
\sigma_{x}U_{0} & = &  \left( \begin{array}{cc}
                                            \vline & \vline \\
                                            \blde_{1} & \blde_{2} \\
                                            \vline & \vline
                                         \end{array} \right) ; \nonumber\\
\sigma_{y}U_{0} & = &  \left( \begin{array}{cc}
                                            \vline & \vline \\
                                            \bldf_{1} & \bldf_{2} \\
                                            \vline & \vline
                                         \end{array} \right) ; \nonumber\\
\sigma_{z}U_{0} & = &  \left( \begin{array}{cc}
                                            \vline & \vline \\
                                            \bldg_{1} & \bldg_{2} \\
                                            \vline & \vline
                                         \end{array} \right) .
\end{eqnarray*}
Then
\begin{subequations}
\label{bigeqn2}
\begin{eqnarray}
\hspace{-0.15in}\barG_{1} = 
  & -\left( \begin{array}{cc}
                  \bldc_{1}^{\dagger}\blde_{1} & \bldc_{1}^{\dagger}\blde_{2} \\
                  \bldc_{2}^{\dagger}\blde_{1} & \bldc_{2}^{\dagger}\blde_{2} 
                \end{array}  \right) 
 & = \left( \begin{array}{cc}
                 \vline & \vline \\
                 \bldgam_{1;1} & \bldgam_{1;2}\\
                 \vline & \vline
             \end{array} \right) ; \\
\hspace{-0.15in}\barG_{2} = 
  & -\left( \begin{array}{cc}
                  \bldc_{1}^{\dagger}\bldf_{1} & \bldc_{1}^{\dagger}\bldf_{2} \\
                  \bldc_{2}^{\dagger}\bldf_{1} & \bldc_{2}^{\dagger}\bldf_{2} 
                \end{array}  \right) 
 & = \left( \begin{array}{cc}
                 \vline & \vline \\
                 \bldgam_{2;1} & \bldgam_{2;2}\\
                 \vline & \vline
             \end{array} \right) ;\\
\hspace{-0.15in}\barG_{3} = 
  & -\left( \begin{array}{cc}
                  \bldc_{1}^{\dagger}\bldg_{1} & \bldc_{1}^{\dagger}\bldg_{2} \\
                  \bldc_{2}^{\dagger}\bldg_{1} & \bldc_{2}^{\dagger}\bldg_{2} 
                \end{array}  \right) 
 & = \left( \begin{array}{cc}
                 \vline & \vline \\
                 \bldgam_{3;1} & \bldgam_{3;2}\\
                 \vline & \vline
             \end{array} \right) ,
\end{eqnarray}
\end{subequations}
which gives
\begin{displaymath}
\bfG_{1} = \left( \begin{array}{c}
                                \bldgam_{1;1}\\
                               \bldgam_{1;2}
                            \end{array} \right) ;
\bfG_{2} = \left( \begin{array}{c}
                                \bldgam_{2;1}\\
                               \bldgam_{2;2}
                            \end{array} \right) ;
\bfG_{3} = \left( \begin{array}{c}
                                \bldgam_{3;1}\\
                               \bldgam_{3;2}
                            \end{array} \right) ,
\end{displaymath}
and
\begin{displaymath}
G = \left( \begin{array}{ccc}
                    \vline & \vline & \vline \\
                    \bfG_{1} & \bfG_{2} & \bfG_{3} \\
                    \vline & \vline & \vline
                \end{array} \right) .
\end{displaymath}
Writing
\begin{displaymath}
\bldw = \left(  \begin{array}{c}
                            w_{1}\\
                            w_{2}\\
                            w_{3}\\
                            w_{4}
                       \end{array} \right)
  = \left(  \begin{array}{c}
                    \bldom_{1}\\
                    \bldom_{2}
               \end{array} \right) 
\end{displaymath}
gives
\begin{equation}
G^{\dagger}\bldw = 
 \left(  \begin{array}{c}
              \bldgam_{1;1}^{\dagger}\bldom_{1} + \bldgam_{1;2}^{\dagger}
                   \bldom_{2}\\
              \bldgam_{2;1}^{\dagger}\bldom_{1} + \bldgam_{2;2}^{\dagger}
                   \bldom_{2}\\
              \bldgam_{3;1}^{\dagger}\bldom_{1} + \bldgam_{3;2}^{\dagger}
                   \bldom_{2}
           \end{array} \right)
=  \left(  \begin{array}{c}
                  \pi_{1}\\
                  \pi_{2}\\
                  \pi_{3}
           \end{array} \right).
\label{bigeqn3}
\end{equation}
With these preliminaries taken care of, we go on to calculate $\calI$.

\subsection{Calculating $\calI$}
\label{appendixB2}

We show how to calculate the matrix element $\calI_{11}$. Calculation of the 
remaining three matrix elements is similar and so we simply quote the final 
result for these matrix elements at the end of this subsection. 

From Eqs.~(\ref{bigeqn1})--(\ref{bigeqn3}) we have
\begin{eqnarray*}
\calI_{11} & = & (\bldc_{1}^{\dagger}\blde_{1})\pi_{1} + 
                             (\bldc_{1}^{\dagger}\bldf_{1})\pi_{2} +
                              (\bldc_{1}^{\dagger}\bldg_{1})\pi_{3} \\
 & = & w_{1}\left[ 
                  ( \bldc_{1}^{\dagger}\blde_{1})(\blde_{1}^{\dagger}\bldc_{1}) 
          +   
                  ( \bldc_{1}^{\dagger}\bldf_{1})(\bldf_{1}^{\dagger}\bldc_{1})
+   
                  ( \bldc_{1}^{\dagger}\bldg_{1})(\bldg_{1}^{\dagger}\bldc_{1})
\right] \\
& & + w_{2}\left[ 
                  ( \bldc_{1}^{\dagger}\blde_{1})(\blde_{1}^{\dagger}\bldc_{2}) 
          +   
                  ( \bldc_{1}^{\dagger}\bldf_{1})(\bldf_{1}^{\dagger}\bldc_{2})
+   
                  ( \bldc_{1}^{\dagger}\bldg_{1})(\bldg_{1}^{\dagger}\bldc_{2})
\right] \\
& & + w_{3}\left[ 
                  ( \bldc_{1}^{\dagger}\blde_{1})(\blde_{2}^{\dagger}\bldc_{1}) 
          +   
                  ( \bldc_{1}^{\dagger}\bldf_{1})(\bldf_{2}^{\dagger}\bldc_{1})
+   
                  ( \bldc_{1}^{\dagger}\bldg_{1})(\bldg_{2}^{\dagger}\bldc_{1})
\right]\\
& & + w_{4}\left[ 
                  ( \bldc_{1}^{\dagger}\blde_{1})(\blde_{2}^{\dagger}\bldc_{2}) 
          +   
                  ( \bldc_{1}^{\dagger}\bldf_{1})(\bldf_{2}^{\dagger}\bldc_{2})
+   
                  ( \bldc_{1}^{\dagger}\bldg_{1})(\bldg_{2}^{\dagger}\bldc_{2})
\right]\\
 & = & Tr\left[ \begin{array}{c}
                           \left(\blde_{1}\blde_{1}^{\dagger} + 
                                     \bldf_{1}\bldf_{1}^{\dagger} +
                                    \bldg_{1}\bldg_{1}^{\dagger}\right) 
                            \left( w_{1}\bldc_{1}\bldc_{1}^{\dagger} +
                                      w_{2}\bldc_{2}\bldc_{1}^{\dagger}\right) \\
                         + \left(\blde_{1}\blde_{2}^{\dagger} + 
                                     \bldf_{1}\bldf_{2}^{\dagger} +
                                    \bldg_{1}\bldg_{2}^{\dagger}\right) 
                            \left( w_{3}\bldc_{1}\bldc_{1}^{\dagger} +
                                      w_{4}\bldc_{2}\bldc_{1}^{\dagger}\right)
                       \end{array} \right].
\end{eqnarray*}
Inserting the various definitions from Appendix~\ref{appendixB1} finally 
gives (after a moderate amount of algebra)
\begin{equation}
\calI_{11} = w_{1} - w_{4} .
\end{equation}
Similar calculations give:
\begin{eqnarray}
\calI_{21} & = & 2w_{2} \\
\calI_{12} & = & 2w_{3} \\
\calI_{22} & = & w_{4}-w_{1}.
\end{eqnarray}
This completes the derivation of Eq.~(\ref{bigsurprise}).

\section{Modeling phase noise effects}
\label{appendixNoiseModel}

In this Appendix we present the noise model used to study the impact of 
phase jitter on the NOC improved TRP gates presented in Section~\ref{sec3b1}. 
Appendix~\ref{appendixC1} introduces the noise model and establishes key 
relations between the noise parameters; while Appendix~\ref{appendixC2} 
describes how a realization of phase noise with arbitrary power is generated, 
as well as the protocol used to simulate the noisy Schrodinger gate dynamics. 

\subsection{Noise model}
\label{appendixC1}

We start with a few basic facts about stationary random processes. 
The rate at which a noise field $N(t)$ can do work (i.~e.\ noise 
power) is \cite{tho},
\begin{displaymath}
P = N^{2}(t)  ,
\end{displaymath}
and the energy that can be delivered in a time interval $dt$ is,
\begin{displaymath}
dE = N^{2}(t)\, dt .
\end{displaymath}
We consider power-type noise for which the time-averaged noise power
\begin{equation}
\label{meanpow}
\overline{P} = \lim_{T\rightarrow\infty}\:\frac{1}{T}\,\int_{-T/2}^{T/2}\,
                 N^{2}(t)\, dt 
\end{equation}
is finite. The total noise energy 
\begin{equation}
\label{Etot}
E = \int_{-\infty}^{\infty}\, dt\, N^{2}(t) 
\end{equation}
diverges for this class of noise. The divergence is due to the 
occurrence of an infinite number of noise fluctuations in the time interval 
$-\infty < t < \infty$. The energy of an individual fluctuation is, however, 
finite. 

The time-averaged noise power $\overline{P}$ can be related to the noise 
correlation function,
\begin{equation}
\label{corfcn}
\overline{N(t)N(t-s)} \equiv \lim_{T\rightarrow\infty}\,\frac{1}{T}\,
                         \int_{-T/2}^{T/2}\, dy\, N(y)N(y-s)  .
\end{equation}
Comparing Eqs.~(\ref{meanpow}) and (\ref{corfcn}) we see that,
\begin{equation}
\label{pow1}
\meanP = \overline{N^{2}(t)}  .
\end{equation}
The Weiner-Khintchine theorem \cite{rie} shows that the noise correlation
function and the power spectral density $S_{N}(f)$ form a Fourier transform
pair: 
\begin{equation}
\label{wkthm}
\overline{N(t)N(t-s)} = \int_{-\infty}^{\infty}\, df\, \specdenN\,
                           e^{-2\pi ifs}  .
\end{equation}
Thus, it follows from Eqs.~(\ref{pow1}) and (\ref{wkthm}) that
\begin{equation}
\label{powspec}
\meanP = \int_{-\infty}^{\infty}\, df\, \specdenN  ,
\end{equation}
which identifies $\specdenN$ as the mean noise power available in the
frequency interval ($f$, $f+df$). 

In the remainder of this Appendix we focus on phase noise $\delta\phi (\tau )$, 
where $\tau$ is the dimensionless time introduced in Appendix~\ref{appendixA1}. 
We model this noise as shot noise which is a common type of electronic noise.
The presentation extends earlier work in Ref.~\cite{fg1}. It is straight-forward
to adapt the following development to treat other forms of noise. 

As shot noise, the phase noise $\delta\phi (\tau )$ is produced by a sequence 
of randomly occurring noise fluctuations $F(t)$. The fluctuations: (1)~occur 
independently of each other at average rate $\meann\,$ per unit time; 
(2)~are uniformly distributed over the time interval $[-\tau_{0}/2,\tau_{0}/2]$
of the TRP inversion; and
(3)~have a peak value $x$ which is Gaussian distributed with mean 
$\overline{x} = 0$, variance $\overline{x^{2}}= \sigma^{2}$, and temporal 
width $2\tau_{f}$ which is the fluctuation lifetime. We assume that
$2\tau_{f}$ is much shorter than the TRP inversion time $\tau_{0}$. The 
bandwidth of $F(\tau )$ is thus $\Delta\omega \sim 1/2\tau_{f}$. Thus a 
realization of the phase noise has the form
\begin{equation}
\delta\phi (\tau ) = \sum_{i=1}^{\mathcal{N}_{f}}\, F(\tau -\tau_{i}) ,
\label{noisedef}
\end{equation}
where $\mathcal{N}_{f}$ denotes the number of noise fluctuations present (a 
stochastic variable), $i$ labels the noise fluctuations, and $\tau_{i}$ specifies 
the center of the $i$th fluctuation. The mean number of fluctuations 
$\overline{\mathcal{N}}_{\hspace{-0.25em}f}$ occurring in the 
time interval $[-\tau_{0}/2,\tau_{0}/2]$ is 
$\overline{\mathcal{N}}_{\hspace{-0.25em}f} = \meann\,  \tau_{0}$. 
It is well-known that for noise with these properties, the actual number of 
fluctuations $n$ that occur in a time $\tau_{0}$ is governed by the Poisson 
distribution \cite{drk}:
\begin{displaymath}
p(n) = \frac{(\overline{\mathcal{N}}_{\hspace{-0.25em}f})^{n}}{n!}\, 
         e^{-\overline{\mathcal{N}}_{\hspace{-0.25em}f}} .
\end{displaymath}
The energy present in a single fluctuation is:
\begin{equation}
\label{flucE}
\varepsilon = \int_{-\infty}^{\infty}\, F^{2}(\tau ) \, d\tau  .
\end{equation}
Let $F(\tau ) = xh(\tau )$, where $h(\tau )$ is any convenient function of 
finite support with normalization
\begin{equation}
\int_{-\infty}^{\infty}\, d\tau\, h^{2}(\tau ) = 2\tau_{f}  .
\label{hsupp}
\end{equation}
As mentioned above, $x$ is Gaussian distributed with mean
$\overline{x} = 0$ and variance $\overline{x^{2}}=\sigma^{2}$. From 
Eq.~(\ref{flucE}), $\varepsilon = 2x^{2}\, \tau_{f}$, and the mean 
energy per fluctuation $\overline{\varepsilon}$ is,
\begin{equation}
\label{meanfE}
\overline{\varepsilon} = 2\,\overline{x^{2}}\,\tau = 2\sigma^{2}\,\tau .
\end{equation}
For shot noise, the power spectral density for $\delta\phi (\tau )$ 
is \cite{ric}
\begin{equation}
\label{campthm}
\specdenphi = \meann\, |g(f)|^{2}  ,
\end{equation}
where $g(f)$ is the Fourier transform of the fluctuation profile
$F(t)$. Thus, using Eqs.~(\ref{powspec}), (\ref{campthm}), and Paresval's
theorem gives,
\begin{equation}
\meanP = \meann\,\int_{-\infty}^{\infty}\, d\tau\, F^{2}(\tau )  .
\end{equation}
Finally, using Eqs.~(\ref{flucE}) and (\ref{meanfE}) gives,
\begin{equation}
\label{finlmeanP}
\meanP = 2\,\meann\,\sigma^{2}\,\tau_{f}  .
\end{equation}
Thus we see that our noise model is characterized by any three of the parameters
$\meanP$, $\meann$, $\sigma^{2}$, and $\tau_{f}$.

We close this subsection by deriving an important connection between the mean 
noise power $\meanP$ and the phase jitter $\sigma_{\phi}$ introduced in 
Section~\ref{sec3c2}. From Eq.~(\ref{noisedef}), we have 
\begin{equation}
\delta\phi^{2}(\tau ) = \sum_{i,j=1}^{\calN_{f}}F(\tau -\tau_{i})
                                                  F(\tau -\tau_{j}) .
\label{noisesq}
\end{equation}
Averaging over the noise gives
\begin{equation}
\overline{\delta\phi^{2}(\tau )} = \barNf \,\overline{F^{2}(\tau )},
\label{noisevar}
\end{equation}
where we have used the statistical independence of distinct noise fluctuations,
and that $2\tau_{f}\ll\tau_{0}$. As in the proof of Campbell's theorem 
\cite{ric2}, it is possible to show that 
\begin{equation}
\overline{F^{2}(\tau )} = \int_{-\infty}^{\infty}\frac{d\tau}{\tau_{0}}\:
                                             \sigma^{2} \, h^{2}(\tau ) ,
\label{campbellconn}
\end{equation}
where, recall $F(\tau ) = x h(\tau )$, and $\overline{x^{2}} = \sigma^{2}$. 
Inserting Eq.~(\ref{campbellconn}) and $\sigma_{\phi}= 
\sqrt{\overline{\delta\phi^{2}(\tau )}}$  into Eq.~(\ref{noisevar}) gives
\begin{equation}
\sigma_{\phi}^{2} = \frac{\barNf}{\tau_{0}} \sigma^{2}\,  
                                      \int_{-\infty}^{\infty} d\tau \:  h^{2}(\tau ) .
\label{tempnoyzvar}
\end{equation}
Finally, inserting Eqs.~(\ref{hsupp}) and (\ref{finlmeanP}), and $\barNf = 
\meann\tau_{0}$ into Eq.~(\ref{tempnoyzvar}) gives  
\begin{equation}
\sigma_{\phi} = \sqrt{\meanP} .
\label{jittpowres}
\end{equation}
Thus the phase jitter $\sigma_{\phi}$ is simply another way to represent the 
phase noise power $\meanP$. Using Eq.~(\ref{jitterconv}), we can also express
the timing jitter $\sigma_{t}$ in terms of $\meanP$: 
\begin{equation}
\sigma_{t} = \frac{\sqrt{\meanP}}{(2\pi f_{clock})}.
\end{equation} 

\subsection{Noisy simulation protocol}
\label{appendixC2}

The numerical simulations used to study the impact of phase jitter on the NOC
improved TRP gates constructs a realization of phase noise as follows. We first 
sample a positive integer $\calN_{f}$ according to the Poisson distribution 
with mean $\overline{\mathcal{N}}_{\hspace{-0.25em}f} = \meann\, \tau_{0}$, 
where $\tau_{0}$ is the (dimensionless) TRP inversion time. $N_{f}$ corresponds 
to the number of fluctuations present in the noise realization. The noise model 
assumes these fluctuations occur independently with probability $dp_{f} = 
(1/\tau_{0})d\tau$. We sample $\calN_{f}$ numbers $\tau_{i}$ ($i= 1, \cdots , 
\calN_{f}$) from the interval $(-\tau_{0}/2,\tau_{0}/2)$. The $\tau_{i}$ give 
the temporal centers of the $\calN_{f}$ fluctuations. For simplicity, we
assume that the fluctuation profile $h(\tau )$ is a square pulse of duration 
$2\tau_{f}$. We next carry out $\calN_{f}$ samples $x_{i}$ ($i = 1, \cdots , 
\calN_{f}$) of a Gaussian distribution with mean $\overline{x}_{i}=0$ and 
variance $\overline{x^{2}_{i}} = \sigma^{2}$. Here $x_{i}$ is the peak 
value of the $i$th fluctuation. These sample results produce the noise 
realization $\delta\Phi (\tau )$:
\begin{equation}
\label{noyzrel}
\delta\Phi (\tau ) = \sum_{i=1}^{\calN_{f}}\, x_{i}\,\left[\,
             \frac{\sgn (\tau - \tau_{il}) - \sgn (\tau-\tau_{ir})}{2}
           \,\right]  , 
\end{equation}
where $\tau_{il} = \tau_{i} -\tau_{f}$, and $\tau_{ir} = \tau_{i}+\tau_{f}$. 
We shall need to produce noise realizations with arbitrary mean noise power 
$\meanP$. We do this by the following normalization procedure. First we 
calculate the mean noise power $\overline{\mathcal{P}}$ of the noise realization
$\delta\Phi (\tau )$ just produced:
\begin{equation}
\label{rawpow}
\overline{\mathcal{P}} = \frac{1}{\tau_{0}}\,\int_{-\tau_{0}/2}^{\tau_{0}/2}\, 
                                            d\tau \, \delta\Phi^{2}(\tau )  .
\end{equation}
Then, if the desired value for the noise power is $\meanP$, we rescale
$\delta\Phi (\tau )$ in Eq.~(\ref{noyzrel}) so that $\delta\Phi (\tau )
\rightarrow \delta\phi (\tau ) \equiv \sqrt{\meanP / \overline{\mathcal{P}}}\, 
\delta\Phi (\tau )$. The result is a noise realization $\delta\phi (\tau )$ with
mean noise power $\meanP$. The simulation takes as inputs the mean noise 
power $\meanP$, the standard deviation $\sqrt{\overline{x_{i}^{2}}} = \sigma$, 
and $\tau_{f}$ which is half the fluctuation lifetime. The fluctuation rate 
$\meann$ then follows from Eq.~(\ref{finlmeanP}): $\meann = \meanP /(2\sigma^{2}
\tau_{f})$. In all the one (two) qubit gate simulations, we used $\sigma = 
0.1$ ($0.1$) and $\tau_{f} = 0.3$ ($0.1$). All one-qubit gates were run at
mean noise power $\meanP = 0.001, 0.008$ corresponding to timing jitter
$\sigma_{t} = 5.03\mathrm{ps}, 14.2\mathrm{ps}$, respectively. The Hadamard gate 
was run at seven other values of $\meanP$ to produce the data displayed in 
Figure~\ref{fig3}. The two-qubit gate was run at $\meanP = 0.001, 0.005$
corresponding to timing jitter $\sigma_{t} = 5.03\mathrm{ps},\; 11.3\mathrm{ps}$.

For a given target gate, and given values of $(\meanP , \sigma , \tau_{f})$,
ten phase noise realizations $\delta\phi (\tau )$ were generated. For each 
realization, the phase noise was added to the TRP twist phase $\phi_{4}(\tau )$,
and the resulting noisy twist phase $\phi^{\prime}_{4}(\tau )$ caused the 
noisy TRP control field $\bff^{\prime}_{0}(\tau )$ to twist incorrectly, as 
described in Section~\ref{sec3c2}. For each noise realization: (i)~the state 
trajectory $U(\tau )$ was determined by numerically simulating the Schrodinger 
dynamics generated by the noisy control field $\bff^{\prime}(\tau ) =
\bff_{0}^{\prime}(\tau ) + \Delta\bff (\tau )$ (see Section~\ref{sec3c2}); 
and (ii)~used to determine the $Tr\, P$ upper bound for the gate error 
probability $P_{e}$. Using the ten values of $Tr\, P$ obtained from the 
simulations, the average $\langle Tr\, P\rangle$ and standard deviation 
$\sigma (TrP)$ were then calculated and the noise-averaged NOC gate 
performance was then approximated by $P_{e}\leq\langle Tr\, P\rangle\pm
\sigma (TrP)$. The results of these simulations appear in 
Section~\ref{sec3c2} and Appendix~\ref{appendixRemainingResults}.\\

\section{Results for remaining quantum gates}
\label{appendixRemainingResults}

In Sections~\ref{sec3b} and \ref{sec3c} we presented our numerical simulation
results for the TRP-NOC improved approximation to the Hadamard gate. In this 
Appendix we present our results for the remaining quantum gates in the universal
gate set $\calGU$ introduced in Section~\ref{sec3a2}. These are the one-qubit
NOT, modified phase, and modified $\pi /8$ gates, and the two-qubit modified 
controlled-phase gate. We present the NOC performance gains for ideal control 
in Appendix~\ref{appendixD1}, and in Appendix~\ref{appendixD2} examine the 
robustness of these gains to: (i)~control parameters with finite precision; and 
(ii)~timing/phase jitter. As our discussion closely follows that in 
Sections~\ref{sec3b} and \ref{sec3c}, a more abbreviated discussion will be
given here. 

\subsection{Ideal control}
\label{appendixD1}

For each one-qubit gate in $\calGU$, the nominal Hamiltonian $H^{1}_{0}(\tau )$ 
(see Eq.~(\ref{oneqbtHam})) is determined by the corresponding values of
$\lambda$ and $\eta_{4}$ appearing in Table~\ref{tableapp1} and the 
dimensionless TRP inversion time $\tau_{0}=160$. With $H^{1}_{0}(\tau )$,
the numerical simulation procedure described in Section~\ref{sec3a3} for 
Strategy~1 was implemented to determine the $Tr\, P$ upper bound on the 
gate error probability $P_{e}\leq Tr\, P$. For the two-qubit modified 
controlled-phase gate, the two-qubit nominal Hamiltonian $H^{2}_{0}(\tau )$ 
(see Eq.~(\ref{twoqbtHam})) is determined by the control parameters appearing 
in Table~\ref{tableapp2} and the dimensionless TRP inversion time $\tau_{0}=
120$. For Strategy~2 , Step~2 of the six step numerical procedure requires the 
three matrices $\calG_{1}$, $\calG_{2}$, and $\calG_{3}$. These follow from 
the functional derivatives of $H^{2}_{0}(\tau )$ with respect to the components 
of the control field $\bff (\tau )$:
\begin{equation}
\label{G2123}
\begin{footnotesize}
\begin{cases}
     \begin{aligned}
     \mathcal{G}_1 = d_3&\left[{\cos{\left({(\frac{d_1 b_2}{b_1 - b_2}+d_1)\tau}
             \right)}\sigma_x^1+\sin{\left({(\frac{d_1 b_2}{b_1 - b_2}+d_1)\tau}
                 \right)}\sigma_y^1}\right]\\
                + & \left[{\cos{\left({(\frac{d_1 b_2}{b_1 - b_2})\tau}\right)}
             \sigma_x^2+\sin{\left({(\frac{d_1 b_2}{b_1 - b_2})\tau}\right)}
                    \sigma_y^2}\right]\\      &  \\
     \mathcal{G}_2 = d_3&\left[{\cos{\left({(\frac{d_1 b_2}{b_1 - b_2}+d_1)\tau}
           \right)}\sigma_y^1-\sin{\left({(\frac{d_1 b_2}{b_1 - b_2}+d_1)\tau}
                  \right)}\sigma_x^1}\right]\\
                 +& \left[{\cos{\left({(\frac{d_1 b_2}{b_1 - b_2})\tau}\right)}
            \sigma_y^2-\sin{\left({(\frac{d_1 b_2}{b_1 - b_2})\tau}\right)}
                \sigma_x^2}\right]\\           &\\
     \mathcal{G}_3 = d_3&\sigma_z^1+\sigma_z^2
     \end{aligned}
     \end{cases}
     \end{footnotesize}
\end{equation}
As noted in Step~3 of the procedure for Strategy~2, we chose $R(\tau ) =
I_{3\times 3}$ and $S(\tau ) = I_{16\times 16}$, where $I_{n\times n}$ is the
$n\times n$ identity matrix. Satisfying the Ricatti equation then required
$Q(\tau ) = G(\tau )G^{\dagger}(\tau )$. Carrying out the remaining steps in the
numerical procedure for Strategy~2 leads to the $Tr\, P$ upper bound for the 
gate error probability $P_{e}$. The simulation results for all gates in the 
universal set $\calGU$ appear in Table~\ref{table1} (see Section~\ref{sec3b1}). 
We see that for all one-qubit 
gates in $\calGU$, NOC reduced the gate error probability $P_{e}$ by 
\textit{four orders-of-magnitude\/} (viz.~$10^{-4}\rightarrow 
10^{-8}$), while for the two-qubit gate, $P_{e}$ was reduced by
\textit{two orders-of-magnitude\/} (viz.~$10^{-3}\rightarrow 10^{-5}$). NOC
has thus substantially improved TRP gate performance, producing gates with 
error probabilities falling well below the target accuracy threshold of 
$10^{-4}$. Because $P_{e}$ is so small for the one-qubit gates, we do not 
write out the unitary matrix produced by NOC as they each agree with their
corresponding target gate $U_{tgt}$ to six significant figures. For the 
two-qubit modified controlled-phase gate, the unitary gate produced is:
\begin{displaymath}
Re(V_{cp}) = \left(   
               \begin{array}{cccc}
                  1.0000 & 0.0001 & 0.0000 & 0.0024 \\
                 0.0000 & 1.0000 & -0.0001 & 0.0000 \\
                  0.0001 & 0.0001 & -1.0000 & -0.0001 \\
                -0.0024 & 0.0000 & 0.0000 & 1.0000\\
              \end{array} \right) ;
\end{displaymath}
\begin{displaymath}
Im(V_{cp}) = \left(   
               \begin{array}{cccc}
                  0.0055 & 0.0001 & 0.0000 & -0.0016 \\
                 -0.0001 & 0.0014 & 0.0004 & 0.0000 \\
                  -0.0001 & -0.0004 & 0.0003 & 0.0000 \\
                -0.0017 & 0.0000 & 0.0000 & 0.0015\\
              \end{array} \right) .
\end{displaymath}
The reader can directly examine the NOC improvement in $V_{cp}$ by comparing 
the above unitary gate with that found in Ref.~\cite{trp6} which was 
reproduced in Appendix~\ref{appendixA3}.

We now determine the amount of bandwidth needed to realize these NOC 
performance improvements. The following calculations assume 
the TRP inversion time for a one-qubit gate is $1\mu\mathrm{s}$ and for the 
two-qubit gate is $5\mu\mathrm{s}$. Recall that the (dimensionless) bandwidth 
was estimated by determining the frequency $\omega_{0.1}$ at which 
$\Delta\calF_{x}(\omega )$ is $10\%$ of the peak value$\Delta\calF_{x}(0)$. For 
the one-qubit gates, Eq.~(\ref{onequbitconvsn}) then determined the dimensionful
bandwidth $\barom_{0.1}$. For the two-qubit gate, whose dimensionless TRP 
inversion time is $\tau_{0} = 120$, the connection between dimensionful and 
dimensionless bandwidth is
\begin{equation}
\frac{\barom_{0.1}}{\omega_{0.1}} = \frac{120}{5\mu\mathrm{s}} = 24\; 
                                                                   \mathrm{MHz}.
\label{twoqubitconvsn}
\end{equation} 

With these preliminaries out of the way, we present our bandwidth results 
for the gates in $\calGU$.\\

\textbf{1.~Hadamard gate:\/} This gate was considered in Section~\ref{sec3b2}.
The (dimensionful) bandwidth found there is $\barom_{0.1} = 640\;
\mathrm{MHz}$.\\

\begin{figure}[!htb]
  \centering
  \includegraphics[trim=1cm 6.5cm 0 6.5cm,clip,
    width=.5\textwidth]{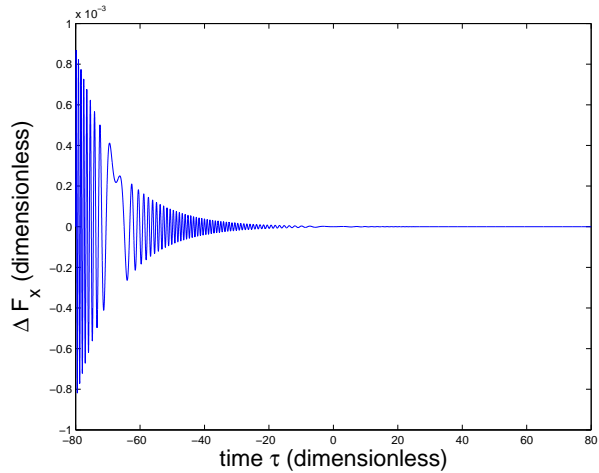}\\
  \caption{(Color online) The $x$-component of the control field modification 
$\Delta F_{x}(\tau )$ for the  NOT gate.}\label{fig4}
\end{figure}

\begin{figure}[!htb]
  \centering
  \includegraphics[trim=1.5cm 6.5cm 0 6cm,clip,
width=.5\textwidth]{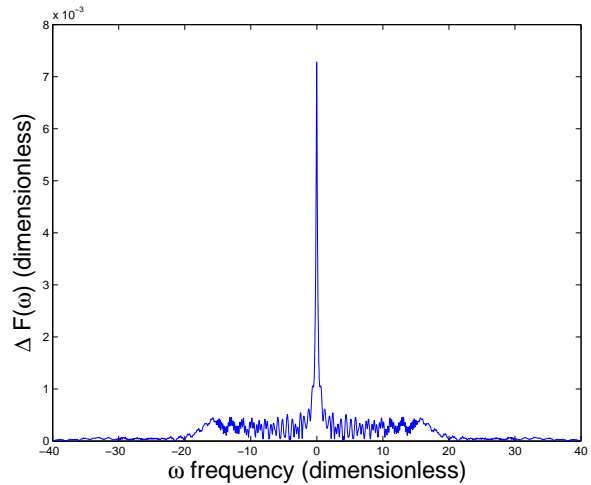}\\
  \caption{(Color online) The Fourier transform of the $x$-component of 
the control field modification $\Delta\calF_{x}(\omega )$ for the NOT 
gate.}\label{fig5}
\end{figure}

\textbf{2.~NOT gate:\/} Figure~\ref{fig4}
shows the x-component of the control field modification $\Delta F_{x}(\tau )$ 
as a function of the dimensionless time $\tau$ for the NOT gate.
Figure~\ref{fig5} 
shows its Fourier transform $\Delta\calF_{x}(\omega )$.
Examination of the data used to produce Figure~\ref{fig5} gives $\omega_{0.1} =
0.8$. Eq.~(\ref{onequbitconvsn}) then gives a dimensionful bandwidth of
$\barom_{0.1} = 130\;\mathrm{MHz}$.\\

\begin{figure}[!htb]
  \centering
  \includegraphics[trim=1cm 6.5cm 0 6cm,clip,
width=.5\textwidth]{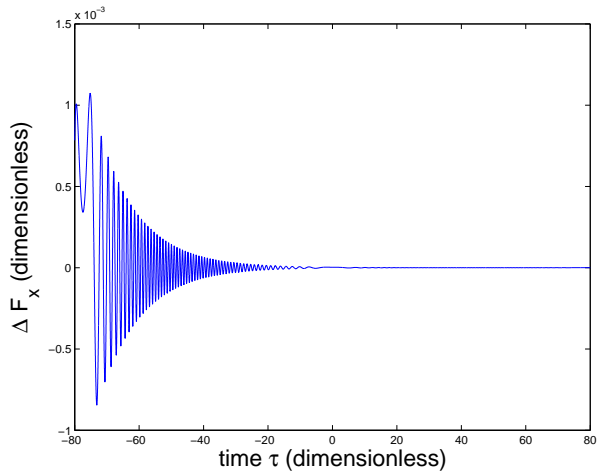}\\
  \caption{(Color online) The $x$-component of the control field modification 
$\Delta F_{x}(\tau )$ for the modified phase gate}\label{fig6}
\end{figure}

\begin{figure}[!htb]
  \centering
  \includegraphics[trim=1cm 6.5cm 0 6cm,clip,
width=.5\textwidth]{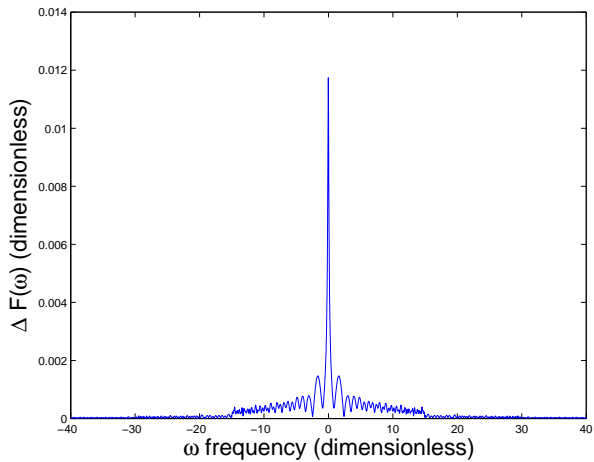}\\
  \caption{(Color online) The Fourier transform of the $x$-component of the 
control field modification $\Delta\calF_{x}(\omega )$ for the modified phase 
gate.}\label{fig7}
\end{figure}

\textbf{3.~Modified phase gate:\/} Figure~\ref{fig6}
shows the x-component of the control field modification $\Delta F_{x}(\tau )$ 
as a function of the dimensionless time $\tau$ for the modified phase gate.
Figure~\ref{fig7} 
shows its Fourier transform $\Delta\calF_{x}(\omega )$. Examination of the data 
used to produce Figure~\ref{fig7} gives $\omega_{0.1} = 1.9$, which, using 
Eq.~(\ref{onequbitconvsn}), gives a dimensionful bandwidth of $\barom_{0.1} = 
300\;\mathrm{MHz}$.\\

\begin{figure}[!htb]
  \centering
  \includegraphics[trim=1cm 6.5cm 0 6cm,clip,
width=.5\textwidth]{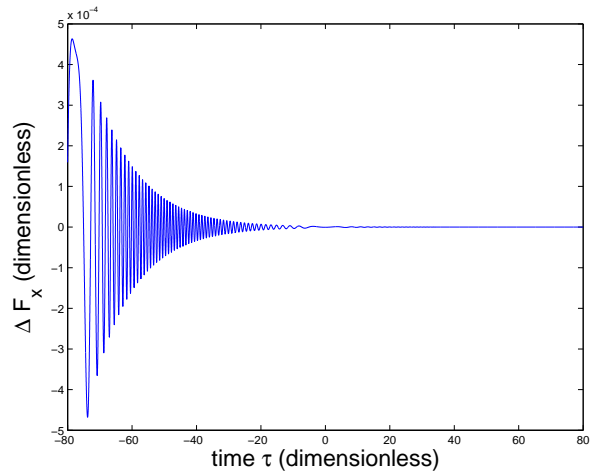}\\
  \caption{(Color online) The $x$-component of the control field modification 
$\Delta F_{x}(\tau )$ for the modified $\pi/8$ gate.}\label{fig8}
\end{figure}

\begin{figure}[!htb]
  \centering
  \includegraphics[trim=1cm 6.5cm 0 6cm,clip,
width=.5\textwidth]{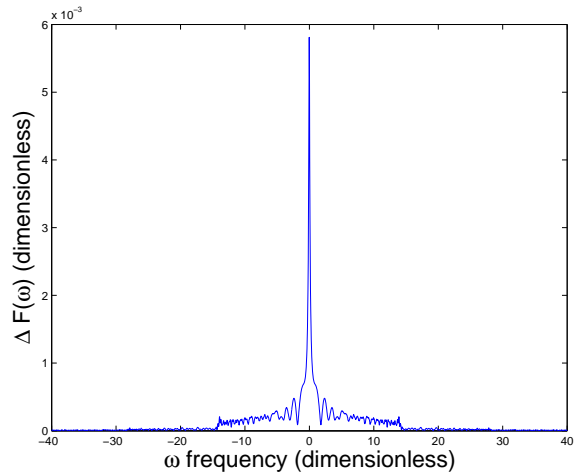}\\
  \caption{(Color online) The Fourier transform of the $x$-component of the 
control field modification $\Delta\calF_{x}(\omega )$ for the modified $\pi/8$ 
gate.}\label{fig9}
\end{figure}

\textbf{4.~Modified $\pi /8$ gate:\/} Figure~\ref{fig8}
shows the x-component of the control field modification $\Delta F_{x}(\tau )$ 
as a function of the dimensionless time $\tau$ for the modified $\pi /8$ gate. 
Figure~\ref{fig9} 
shows its Fourier transform $\Delta\calF_{x}(\omega )$. Examination of the data 
used to produce Figure~\ref{fig9} gives $\omega_{0.1} = 1.3$, which, using 
Eq.~(\ref{onequbitconvsn}), gives a dimensionful bandwidth of $\barom_{0.1} = 
210\;\mathrm{MHz}$.\\

\begin{figure}[!htb]
  \centering
  \includegraphics[trim=1cm 6.5cm 0 6cm,clip,
width=.5\textwidth]{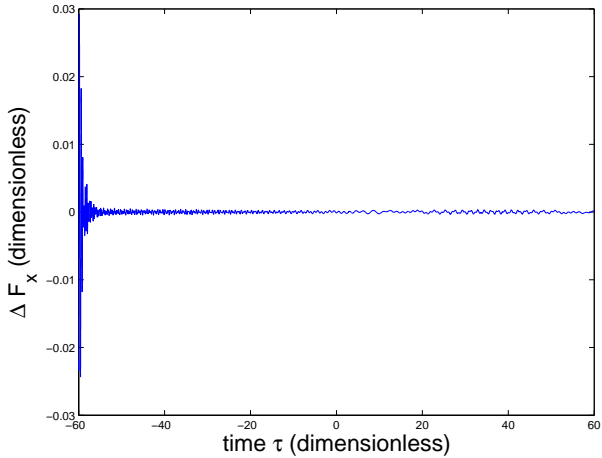}\\
  \caption{(Color online) The $x$-component of the control field modification 
$\Delta F_{x}(\tau )$ for modified controlled-phase gate,}\label{fig10}
\end{figure}

\begin{figure}[!htb]
  \centering
  \includegraphics[trim=1cm 6.5cm 0 6cm,clip,
width=.5\textwidth]{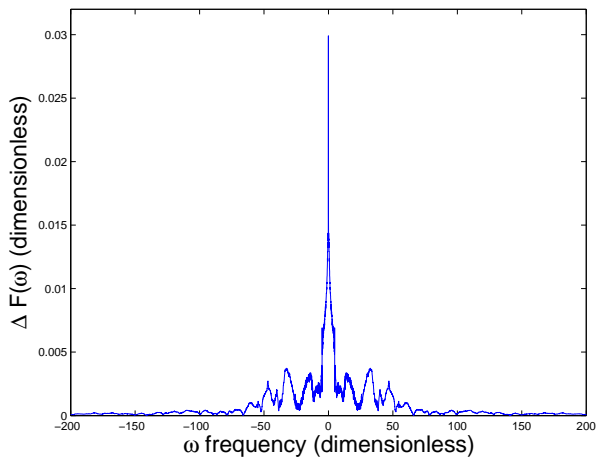}\\
  \caption{(Color online) The Fourier transform of the $x$-component of the 
control field modification $\Delta\calF_{x}(\omega )$ for the modified 
controlled-phase gate.}\label{fig11}
\end{figure}

\textbf{5.~Modified controlled-phase gate:\/} Figure~\ref{fig10}
shows the x-component of the control field modification $\Delta F_{x}(\tau )$ 
as a function of the dimensionless time $\tau$ for the modified controlled-phase
gate. Figure~\ref{fig11} 
shows its Fourier transform $\Delta\calF_{x}(\omega )$. Examination of the data 
used to produce Figure~\ref{fig11} gives $\omega_{0.1} = 34$, which, using 
Eq.~(\ref{twoqubitconvsn}), gives a dimensionful bandwidth of $\barom_{0.1} = 
820\;\mathrm{MHz}$.\\

Table~\ref{table2} (see Section~\ref{sec3b2}) collects the bandwidth results 
for all gates in $\calGU$. As noted there, AWGs with $5\mathrm{GHz}$  
bandwidth are commercially available so that the bandwidth requirements
for NOC are within the range of existing commercially available AWGs.

\subsection{Robustness to imperfect control}
\label{appendixD2}

In this subsection we examine the robustness of the non-Hadamard gates in 
$\calGU$ to: (i)~control parameters with finite-precision 
(Appendix~\ref{appendixD2a}); and (ii)~phase/timing jitter 
(Appendix~\ref{appendixD2b}). The same issues were examined for the Hadamard
gate in Section~\ref{sec3c2}.

\subsubsection{Finite-precision control parameters}
\label{appendixD2a}

As with the discussion of the Hadamard gate in Section~\ref{sec3c1}, here we 
determine the minimum control parameter precision needed to realize the NOC 
performance improvements found for the non-Hadamard gates in $\calGU$
in Appendix~\ref{appendixD1}. For the one-qubit gates, the NOC performance
improvements were found to be most sensitive to small changes in $\eta_{4}$.
Thus we will only show how the $Tr\, P$ upper bound on the gate error
probability $P_{e}$ varied as we changed $\eta_{4}$ by one in its least 
significant digit. For the two-qubit modified controlled-phase gate, performance 
was most sensitive to small changes in $d_{1}$, $d_{4}$, and $c_{4}$. We only 
show results for $d_{1}$ as similar results are found for $d_{4}$ and $c_{4}$.\\

\textbf{1.~NOT gate:\/} For the NOT gate, NOC delivered a gate with $P_{e}
\leq 8.58\times 10^{-9}$. In Table~\ref{tableapp3}
\begin{table}[!htbp]
\centering
 \caption{\label{tableapp3}Sensitivity of $TrP$ to a small variation of $\eta_4$ 
away from its optimum value for the one-qubit NOT gate. For all 
$\eta_{4}$ values, $\lambda$ is maintained at its optimum value $\lambda = 
6.965$. Column~2 (3) shows the variation of $Tr\, P$ when the control 
field includes (omits) the NOC modification $\Delta\bff(\tau )$. Recall 
that $Tr\, P$ upper bounds the gate error probability $P_{e}\leq Tr\, P$.\\}
  \begin{ruledtabular}
\begin{tabular}{ccc}
$\eta_4$ & \hspace{0.22in}$TrP$ (with NOC) & \hspace{0.22in}$TrP$ (without 
NOC)   \\   \hline
 $2.188\times 10^{-4}$ & $6.50\times 10^{-3}$ & $1.55\times 10^{-2}$ \\
 $2.189\times 10^{-4}$ & $8.58\times 10^{-9}$ & $6.27\times 10^{-5}$ \\
  $2.190\times 10^{-4}$ & $9.80\times 10^{-3}$ & $3.28\times 10^{-2}$ \\
\end{tabular}
\end{ruledtabular}
\end{table}
we show how the $Tr\, P$ upper bound on the gate error probability 
($P_{e}\leq Tr\, P$) changes due to a small shift in $\eta_{4}$ away from its 
optimum value. We show the variation in $Tr\, P$ when the NOC modification
is both included and omitted. As with the Hadamard gate, $\eta_{4}$ must be 
controlled to better than one part in $10,000$ to realize the NOC performance 
gains. As shown in the Hadamard gate discussion, this is possible using an AWG 
with at least $14$-bit vertical resolution. Using less precision will give rise to 
uncertainty in the fourth significant digit, and to a washing out of the NOC 
performance gains.\\

\textbf{2.~Modified $\pi /8$ gate:\/} For the modified $\pi /8$ gate, NOC 
delivered a gate with $P_{e}\leq 1.06\times 10^{-8}$. In Table~\ref{tableapp4}
\begin{table}[!htbp]
\centering
 \caption{\label{tableapp4}Sensitivity of $TrP$ to a small variation of $\eta_4$ 
away from its optimum value for the one-qubit modified $\pi /8$ gate. For all 
$\eta_{4}$ values, $\lambda$ is maintained at its optimum value $\lambda = 
8.465$. Column~2 (3) shows the variation of $Tr\, P$ when the control 
field includes (omits) the NOC modification $\Delta\bff(\tau )$. Recall 
that $Tr\, P$ upper bounds the gate error probability $P_{e}\leq Tr\, P$.\\}
  \begin{ruledtabular}
\begin{tabular}{ccc}
$\eta_4$ & \hspace{0.22in}$TrP$ (with NOC) & \hspace{0.22in}$TrP$ (without 
NOC)   \\   \hline
 $1.674\times 10^{-4}$ & $7.10\times 10^{-3}$ & $4.99\times 10^{-2}$ \\
 $1.675\times 10^{-4}$ & $1.06\times 10^{-8}$ & $2.13\times 10^{-4}$ \\
  $1.676\times 10^{-4}$ & $7.30\times 10^{-3}$ & $3.90\times 10^{-2}$ \\
\end{tabular}
\end{ruledtabular}
\end{table}
we show how the $Tr\, P$ upper bound on the gate error probability 
($P_{e}\leq Tr\, P$) changes due to a small shift in $\eta_{4}$ away from its 
optimum value. We show the variation in $Tr\, P$ when the NOC modification
is both included and omitted. As with the Hadamard gate, 
$\eta_{4}$ must be controlled to better than one part in $10,000$ to realize
the NOC performance gains. This is possible using an AWG with at least 
$14$-bit vertical resolution. Using less precision will give rise to uncertainty
in the fourth significant digit, and to a washing out of the NOC performance
gains.\\

\textbf{3.~Modified phase gate:\/} For the modified phase gate, NOC 
delivered a gate with $P_{e}\leq 1.08\times 10^{-8}$. In Table~\ref{tableapp5}
\begin{table}[!htbp]
\centering
 \caption{\label{tableapp5}Sensitivity of $TrP$ to a small variation of $\eta_4$ 
away from its optimum value for the one-qubit modified phase gate. For all 
$\eta_{4}$ values, $\lambda$ is maintained at its optimum value $\lambda = 
8.073$. Column~2 (3) shows the variation of $Tr\, P$ when the control 
field includes (omits) the NOC modification $\Delta\bff(\tau )$. Recall 
that $Tr\, P$ upper bounds the gate error probability $P_{e}\leq Tr\, P$.\\}
  \begin{ruledtabular}
\begin{tabular}{ccc}
$\eta_4$ & \hspace{0.22in}$TrP$ (with NOC) & \hspace{0.22in}$TrP$ (without 
NOC)   \\   \hline
 $1.665\times 10^{-4}$ & $1.20\times 10^{-3}$ & $4.42\times 10^{-2}$ \\
 $1.666\times 10^{-4}$ & $1.08\times 10^{-8}$ & $4.62\times 10^{-4}$ \\
  $1.667\times 10^{-4}$ & $6.10\times 10^{-3}$ & $5.74\times 10^{-2}$ \\
\end{tabular}
\end{ruledtabular}
\end{table}
we show how the $Tr\, P$ upper bound on the gate error probability 
($P_{e}\leq Tr\, P$) changes due to a small shift in $\eta_{4}$ away from its 
optimum value. We show the variation in $Tr\, P$ when the NOC modification
is both included and omitted. As with the Hadamard gate, 
$\eta_{4}$ must be controlled to better than one part in $10,000$ to realize
the NOC performance gains. This is possible using an AWG with at least 
$14$-bit vertical resolution. Using less precision will give rise to uncertainty
in the fourth significant digit, and to a washing out of the NOC performance
gains.\\

\textbf{4.~Modified controlled-phase gate:\/} For the two-qubit modified 
controlled-phase gate, NOC delivered a gate with $P_{e}\leq 5.21\times 
10^{-5}$. In Table~\ref{tableapp6}
\begin{table}[!htbp]
\centering
 \caption{\label{tableapp6}Sensitivity of $TrP$ to a small variation of $d_{1}$ 
away from its optimum value for the two-qubit modified controlled-phase gate. 
For all $d_{1}$ values, the remaining control parameters appearing in 
Table~\ref{tableapp2} are maintained at the optimum values given there. 
Column~2 (3) shows the variation of $Tr\, P$ when the control 
field includes (omits) the NOC modification $\Delta\bff(\tau )$. Recall 
that $Tr\, P$ upper bounds the gate error probability $P_{e}\leq Tr\, P$.\\}
  \begin{ruledtabular}
\begin{tabular}{ccc}
$d_{1}$ & \hspace{0.22in}$TrP$ (with NOC) & \hspace{0.22in}$TrP$ (without 
NOC)   \\   \hline
 $11.701$ & $1.16\times 10^{-3}$ & $3.36\times 10^{-3}$ \\
 $11.702$ & $5.21\times 10^{-5}$ & $1.27\times 10^{-3}$ \\
  $11.703$ & $1.16\times 10^{-3}$ & $1.43\times 10^{-3}$ \\
\end{tabular}
\end{ruledtabular}
\end{table}
we show how the $Tr\, P$ upper bound on the gate error probability 
($P_{e}\leq Tr\, P$) changes due to a small shift in $d_{1}$ away from its 
optimum value. We show the variation in $Tr\, P$ when the NOC modification
is both included and omitted. We see that $d_{1}$ must be 
controlled to better than one part in $100,000$ to realize the NOC performance 
gains. Such control parameter precision is attainable using an AWG with $17$-bit 
vertical resolution (viz.~one part in $2^{17} = 131,072$). We are not aware
of such AWGs being commercially available, thus requiring custom electronics to
realize the NOC performance gains for this two-qubit gate. Note that $16$-bit 
precision corresponds to a precision of one part in $2^{16} = 65,536$, and so 
to an uncertainty in the fifth significant digit. Thus with less than $17$-bits of 
precision, Table~\ref{tableapp6} indicates that the NOC performance gains 
will be washed out by the uncertainty in the least significant digit of $d_{1}$.
Similar results are found for $d_{4}$ and $c_{4}$.

\subsubsection{Phase/timing jitter}
\label{appendixD2b}

In Section~\ref{sec3c2} we discussed the effects of timing/phase jitter on the 
NOC performance gains shown in Table~\ref{table1} of Section~\ref{sec3b1}.
Appendix~\ref{appendixNoiseModel} introduced our model for phase noise and
detailed the protocol for the numerical simulation of the NOC gate dynamics in 
the presence of such noise. Table~\ref{table5} presented the simulation results 
for all gates in $\calGU$ for timing jitter $\sigma_{t}=5\mathrm{ps}$, the same 
as found in commercially available AWGs \cite{jitter}. The Hadamard gate 
was discussed in Section~\ref{sec3c2} and similar remarks apply to the other 
gates in $\calGU$. The noise power corresponding to $5\mathrm{ps}$ timing 
jitter at a clock frequency $f_{clock} =1\mathrm{GHz}$ is $\meanP = 0.001$. 
As discussed in Appendix~\ref{appendixC2}, the one-qubit simulations used
noise fluctuation parameters $\sigma = 0.1$ and $\tau_{f} = 0.3$, while the
two-qubit simulations used $\sigma = 0.1$ and $\tau_{f} = 0.1$. From 
Appendix~\ref{appendixC1}, this corresponds to an average noise fluctuation 
rate $\meann = \meanP /(2\sigma^{2}\tau_{f}) = 0.167\; (0.500)$ for the 
one-qubit (two-qubit) gate simulations. Thus for the one-qubit (two-qubit) 
gates with TRP (dimensionless) inversion time $\tau_{0} = 160\; (120)$, each 
phase noise realization contained, on average, $\calN_{f} =27\; (60)$ noise 
fluctuations.

In Table~\ref{tableapp7}
\begin{table*}[!htb]
\centering
\caption{\label{tableapp7}Sensitivity of $TrP$ to timing jitter $\sigma_{t} = 
\sqrt{\meanP}/(2\pi f_{clock})$ for all target gates in the 
universal set $\mathcal{G}_U$. For all one-qubit (two-qubit) gates, the 
numerical simulations used mean noise power $\bar{P}=0.008\; (0.005)$, 
which corresponds to timing jitter $\sigma_{t}=14.2\; (11.3) \mathrm{ps}$ 
for $f_{clock} = 1\mathrm{GHz}$. For each gate, ten phase noise realizations 
were generated (see Appendix~\ref{appendixNoiseModel}), leading to ten 
values of the $Tr\, P$ upper bound on the gate error probability $P_{e}\leq 
Tr\, P$. The third column lists, for each gate, the corresponding average 
$<TrP>$, and uses the standard deviation $\sigma (TrP)$ to indicate the 
spread of $Tr\, P$ about the average.\\}
\begin{ruledtabular}
\begin{tabular}{ccc}

  Gate&            Timing-jitter $\sigma_{t}$  & $P_{e}\leq \;\; <TrP> \pm \sigma 
\mathrm{(TrP)}$ with NOC \\
  \hline
                         Hadamard & $14.2ps$ & $(5.58\pm 2.55)\times 10^{-5}$\\
                                  NOT & $14.2ps$ & $(5.71\pm 2.67)\times 10^{-5}$\\  
                 Modified phase & $14.2ps$ & $(7.09\pm 3.23)\times 10^{-5}$\\
               Modified $\pi/8$ & $14.2ps$ & $(8.04\pm 2.43)\times 10^{-5}$\\
Modified controlled phase & $11.3ps$ & $(6.74\pm 1.09)\times 10^{-5}$\\
\end{tabular}
\end{ruledtabular}
\end{table*}
we present further noisy simulation results for all gates in $\calGU$ at 
noise power $\meanP = 0.005\; (0.008)$ for the two-qubit (one-qubit) gate(s). 
This corresponds, respectively, to: (i)~timing jitter $\sigma_{t} = 11.3\; 
(14.2)\mathrm{ps}$; (ii)~$\meann = 2.50\; (1.33)$; and (iii)~phase noise 
realizations with, on average, $\calN_{f} = 300\; (213)$ noise fluctuations. 
We see that the increased noise power $\meanP = 0.001 \rightarrow 0.005, 
0.008$ only degraded the NOC performance gains slightly more than was  
seen in Table~\ref{table5}. Notice that, even with phase jitter that is worse 
than occurs in commercially available AWGs, all gates in $\calGU$ still have 
error probabilities that fall below the target accuracy threshold of $10^{-4}$.


\begin{thebibliography}{99}
\bibitem{foot1} We will spell out what we mean by ``good approximation''
in Section~\ref{sec3}.
\bibitem{foot2} Value functions can also be considered with optimal assignments 
producing maximum value.
\bibitem{foot3} For our purposes, it is sufficient that $Q(t)$ be a positive 
Hermitian matrix.
\bibitem{foot6} As will be discussed in Section~\ref{sec3a3} and
Appendix~\ref{appendixA1}, the numerical simulations transform the dimensionful 
Schrodinger equation into dimensionless form so that the time parameter in the 
simulations is dimensionless. When discussing Strategy~1 in Section~\ref{sec2d} 
the time $t$ is actually dimensionless time. Thus when we say that the one-qubit
numerical simulations have $T=160$ we are making a statement about the 
dimensionless time parameter. The connection between dimensionful and 
dimensionless parameters is explained in Section~\ref{sec3a3} and 
Appendix~\ref{appendixA1}.
\bibitem{foot4} To achieve this level of gate performance required TRP sweep 
parameter values with $17$-bits of precision. Ref.~\cite{trp7} modified the 
optimization cost function to reward both gate robustness and low error 
probability, and found TRP sweep parameter values that produced one-qubit 
TRP gates with $P_{e} < 4.62\times 10^{-4}$, while only requiring sweep 
parameters with $14$-bits of precision. This approach improved gate robustness
at the cost of a small increase in the gate error probability. By showing how 
to inter-leave a TRP sweep with a suitable dynamical decoupling sequence, 
Ref.~\cite{trp8} was able to reduce the gate error probability for the 
two-qubit modified controlled phase gate to $P_{e}< 8.87\times 10^{-5}$, 
while only requiring TRP sweep parameters with $14$-bits of precision. This 
improved performance was achieved at the cost of a more complicated control 
field.
\bibitem{errprobdef} The accuracy threshold theorem (ATT) requires $P_{e}<
P_{a}$ for every application of a quantum gate. By defining $P_{e}$ to be the 
worst-case value of $P_{e}(\psi )$ we insure that if $P_{e}<P_{a}$, then for 
every initial state $\psi$ (and thus for every gate application), $P_{e}(\psi )
<P_{a}$ as demanded by the ATT. However, if $P_{e}$ is defined as an 
average-case value, states could exist for which $P_{e}(\psi ) >P_{a}$ even 
though $P_{e}<P_{a}$. Such states would violate a core requirement of the ATT,
and thus nullify its conclusion. It is not enough to satisfy the conditions of
the ATT on average. Thus, in the context of fault-tolerant quantum computing,
it is more appropriate to define the gate error probability as a worst-case 
value rather than as an average-case value.
\bibitem{jitter} For example, the Tektronix AWG5000 arbitrary waveform generator
has $5\mathrm{ps}$ timing jitter.
\bibitem{foot5} Note that Refs.~\cite{trp5}--\cite{trp8} have a factor of two
error in the quoted value of $\tau_{0}$ for the one-qubit gates. The correct 
value for these gates is $\tau_{0} = 160$, as given here.
\end{thebibliography}
\end{document}